\documentstyle[12pt]{article}
\begin{document}
\title{Scaling of Particle Trajectories on a Lattice I: Critical
 Behavior}
\author{ Meng-She Cao and E.G.D. Cohen \\
The Rockefeller University \\
1230 York Avenue, New York, NY 10021}
\maketitle
\vspace{0.3in}
\begin{center} \bf Abstract \end{center}

The scaling behavior of the closed trajectories of a moving 
particle generated by randomly placed rotators or mirrors on a square
or triangular lattice is studied numerically. For most 
concentrations of the scatterers the trajectories close exponentially 
fast. For special critical concentrations infinitely extended 
trajectories can occur which exhibit a scaling behavior similar to 
that of the perimeters of percolation clusters. In addition to 
the two critical exponents $\tau=15/7$ and $d_{f}=7/4$ found before,
the critical exponent $\sigma=3/7$, which is associated with
the scaling function for trajectory size away from criticality,
also appears. This exponent determines structural scaling 
properties of closed trajectories of finite size when they approach 
infinity, at criticality. New scaling behavior was found for the 
square lattice partially occupied by rotators, indicating a 
different universality class from that of percolation clusters. An argument
for the scaling behavior found along the critical lines is presented.

\vspace{1.0cm}

\newpage
\pagestyle{plain}
\baselineskip=1.5\baselineskip

\section{Introduction}
In a number of previous publications$^{(1-10)}$, the diffusion 
properties of Lorentz lattice gas cellular automata (LLGCA) has 
been studied. There the behavior of a point particle moving through
fixed, regularly$^{(6)}$ or randomly$^{(1-5,7-10)}$ placed 
scatterers on the lattice sites of a variety of planar lattices has
been studied numerically. The scatterers consisted either of 
reflecting right and left mirrors or of rotating right and left   
rotators, which scatter the particle either to its right or its
left, respectively (Fig. 1-2). The particle is constrained to 
move in unit time steps along the lattice bonds (of unit length)
and the lattices studied were the square$^{(1,2,4,7)}$, triangular
$^{(3,7)}$, honeycomb$^{(8)}$, quasi-lattice$^{(8)}$ and random
lattice$^{(3,9)}$. Since these scatterers are fixed on the lattice,
the particle simply travels periodically in the plane once
the trajectory forms a closed orbit. In our computer simulations, 
the particle is stopped right after the first period. In almost
all cases studied so far, all the trajectories close exponentially fast. 
However, in some cases they close power law slow. 
While in the former case all orbits appear to close after typically
$2^{10}$ time steps, in the latter case there are extented 
trajectories, possibly of infinite length, which only close after 
a long (possibly infinite) time. Studying these extented trajectories 
reveals that they exhibit scaling properties, which in the
case of a lattice fully occupied  by scatterers, can be mapped 
onto a corresponding bond or site percolation problem. However, 
in the case of a lattice not fully occupied by scatterers, where
empty sites occur and the particle trajectory can cross itself,
no such mapping seems possible. Nevertheless some of the scaling
properties of these trajectories are then still the same as 
those found for the fully occupied lattice where no such crossing 
can occur(Fig. 3). The identity of the scaling properties of the
closed trajectories with those of a corresponding percolation 
problem is exemplified, for example, by the appearance of the same
two universal critical exponents that occur in the two dimensional
(bond or site) percolation problem: a fractal dimension 
$d_{f}=7/4$ associated with the length $S$ of the trajectories 
(perimeter of the percolation cluster) and the mean square distance 
of all points on a large trajectory of length $S$ from the origin,
$R_S^2$,
\begin{eqnarray}
R_S^2 \sim S^{2/d_{f}}
\end{eqnarray}
as well as the Fisher exponent $\tau=15/7$, characterizing the 
probability distribution of closed trajectories of a certain length,
\begin{eqnarray}
n_{S} \sim S^{-\tau+1}
\end{eqnarray}
A hyperscaling relation holds between these exponents:
\begin{eqnarray}
\tau-1= \frac{2}{d_f} \label{rel}
\end{eqnarray}

We note that the closed particle trajectories, when mapped onto 
percolation clusters, are characterized here by their ``surface'
' properties, i.e, their length, not, as is usually done, by 
their ``bulk '' properties, i.e, the total number of lattice sites 
they contain. The exact values for $d_f$ were first conjectured 
to be $4/7$ by Saproval, Rosso, and Gouyet$^{(11)}$. Ziff$^{(12)}$, 
then, developed a scaling theory which implied that $\tau$ 
should be $15/7$ and he verified numerically the 
values of $d_f$ and $\tau$. Later the conjecture was proved by Saleur
and Duplantier$^{(9)}$ by mapping the percolation problem onto
a Coulomb gas. However, for the partially occupied lattice at 
criticality, there is no theoretical explanation for the 
universality of $\tau$ and $d_f$. 

In this and the following paper, we introduce finer characterizations 
of the closed particle trajectories than we have done in previous 
publications, where only the trajectory size distribution scaling 
exponent $\tau$ and the trajectory's fractal dimension
$d_f$ were considered. The quantities that describe the finer 
characterizations are: (1) The number of right, left scatterers 
and empty sites on a closed trajectory, i.e. $N_R$, $N_L$ and $
N_E$, respectively; (2) The winding angle $W$, i.e. the number 
of right turns minus the number of left turns of the particle moving
on a closed trajectory; (3) The frequency with which lattice
sites on the particle trajectory are visited by the moving particle. 
These are more detailed ``structural'' properties than the
``gross'' properties incorporated in $\tau$ and $d_f$. By studying
these properties for finite trajectories of increasing length
at criticality, we derive, in this paper, a number of trajectory
scaling properties which include not only the asymptotic behavior
for infinitely large closed trajectories, but also the approach
to the asymptotic behavior. As an example of the new properties
that we have just introduced, we quote one that occurs on 
the fully occupied square lattice, where each site is occupied 
either by a right or by a left rotator. If $N_{R}$ and $N_{L}$ are
the numbers of right rotators and  left rotators contained in a
closed trajectory respectively, asymptotically $N_R$ and $N_L$ 
satisfy:
\begin{eqnarray}
<\frac{N_{R}}{N}- \frac{1}{2}>_c=<\frac{N_{L}}{N}- \frac{1}{2}>_
{cc} \sim N^{-0.57} \label{ch1q2}
\end{eqnarray}
where $N=N_{R}+N_{L}$ is the total number of sites of the closed
trajectory. The averages $<\cdots>_c$ and $<\cdots>_{cc}$ are 
taken over all closed trajectories containing $N$ sites in which 
the particle moves clockwise (Fig. 3) and counter-clockwise, 
respectively. There are two observations to  make with regards to  
Eq.~(\ref{ch1q2}). First, the number of right and left rotators on
a closed orbit become asymptotically equal, i.e. asymptotically
$N_{R}/N=N_{L}/N=1/2$. Second the critical exponent $0.57$ is 
in good approximation equal to $1-\sigma$, where the critical
cluster perimeter length exponent $\sigma=3/7$. All scaling results
observed so far for the fully occupied square as well as 
triangular lattice are of the form of Eq.(4), where the constant 
inside the average (here equal to the critical concentration 
$C_R=C_L=1/2$) can vary, while for the partially occupied 
lattice the exponent $0.57$ can change to its mean field value $0
.50$. In the context of Eq.~(\ref{ch1q2}), the difference of $0.
57$ and $0.50$ means that asymptotically the right and left rotators are
almost but not quite randomly distributed over the trajectory, which 
is not surprising, since some correlations between
the placing of right and left rotators on the lattice sites to
generate a closed trajectory, would be expected. Surprising, 
though, is the universality of the exponent of 0.57, for which we 
have, at present, no explanation. 

Another interesting feature is that some sites on the trajectories
are only visited once by the moving particle and others can be
visited more than once. The number of sites visited by the moving
particle a different number of times exhibits a power law 
behavior similar to Eq.~(\ref{ch1q2}), except for the square lattice
partially occupied by rotator scatterers, where the scaling 
behavior differs significantly from all other cases.

The numerical algorithm we used was obtained from an efficient 
combination of the Ziff algorithm$^{(4,12,14)}$ (use of a virtual 
lattice) and a technique recently developed by Wang and Cohen$^{(7,8)}$ 
(dynamical memory allocation). The simulation 
was done on a virtual lattice of size $65536 \times 65536$.
The lattice was divided into $1024 \times 1024$ blocks of $64 
\times 64$ sites rather than $256 \times 256$ blocks of 
$256 \times 256$ sites as used by Ziff$^{(4)}$ and Wang and Cohen$^{(7,8)}$, 
since a smaller block size is more efficient for dilute scatterer
models. $16$ bits ($2^{16}=65536$) were used to determine the 
coordinates $(x,y)$ of a site, the upper $10$ bits ($2^{10}=1024
$) were used for the location of the block which had been visited
by the moving particle, while the lower $6$ bits ($2^6=64$) were
used for the site position within the block. Another array of
$1024 \times 1024$ was introduced, to record whether a block had
been visited by the particle. The application of bit shifting,
masking etc. to look up the coordinates of blocks and sites 
contained in the blocks made the whole process very fast. Furthermore,
we used a dynamic memory allocation technique, where an array
of pointers is generated so that each block has a corresponding
pointer and actual memory of the states of the sites (i.e. the
type of scatterers placed on the site) in the block is not assigned
to its pointer until the particle enters it (using MALLOC in
C). After a trajectory is finished, only memories that have been
assigned to the pointers are deleted (using FREE in C).

The advantage of this scheme is that since large trajectories are
essentially fractals with fractal dimension $d_f=7/4$, only a 
small fraction of the memory for a $65536 \times 65536$ array is
 actually used and we do not need to reserve memory for those 
areas that are never visited by the particle. Note that since the
 fluctuations in size of the trajectories are very big, our dynamic
memory allocation technique saves an enormous amount of memory
. For example, only a small array is generated for a small 
trajectory no matter how large other trajectories are. After the 
trajectory is completed, only the blocks that have been entered by
 the trajectories need to be reset to the blank condition so that
another independent particle can be launched, rather than 
resetting the entire lattice. Therefore the average time required for
this operation is significantly reduced. Since the memory is 
allocated dynamically, we do not need to know how much memory the
 largest trajectory requires before we start our numerical simulations.

About $300,000$ independent particles, initially placed randomly
 on the lattice, were studied. We only collect closed particle 
trajectories whose lengths are smaller than a certain limit. The 
trajectory is disregarded if it did not close by that number of 
steps. The value of the limit we used was up to $2^{21}-2^{24}$ 
time steps, depending on the concentration of the scatterers. In
 all the simulations, there was no particle that crossed the 
boundary of the virtual lattice.

We have verified for the fully occupied square lattice that our
calculations have been made for systems of a sufficiently large number 
of particles and for sufficiently long times that the error bars in our figures
are typically of the order of the size of the symbols.  The same obtains
for the fully occupied triangular lattice.  However, for the partially
occupied square and triangular lattices the accuracy of our calculations 
decreases with the concentration of scatterers. 

The organization of this paper is as follows. In section 2, we 
introduce the rotator model on the square lattice and discuss the
 closed trajectory scaling results for the rotator model both for
 a fully and for a partially occupied lattice. In section 3, the
 same is done for the mirror model. In section 4, we discuss the
 closed trajectory scaling results for the triangular lattice, 
which are the same for both the rotator and the mirror model, 
since they can be mapped in to each other for this lattice $^{(3,7
)}$. Section 5 contains a summary of our results in three tables
 as well as a discussion and an outlook. In the following paper 
scaling results, in particular the scaling function, describing 
the scaling behavior in the critical region near criticality will
 be given. The same exponent $\sigma$ that played a dominant 
role in this paper at criticality, will also appear there.
\section{Critical behavior of the rotator model on the square 
lattice}
\subsection{The fully occupied lattice}
The rotator model is a special case of a general class of models
, originally introduced by Gunn and Ortu\~{n}o$^{(15)}$. Fixed 
right (Fig. 1a) and left rotators (Fig. 1b) are randomly placed on
 the sites of the lattice. A particle moves along the bonds of 
the lattice and its velocity is rotated either to its right or 
to its left by $\pi/2$ upon being scattered by a right or a left 
rotator, respectively (Fig. 3a). The total concentration of the 
scatterers $C$ is the sum of those of the right rotators $C_R$ 
and the left rotators $C_L$: $C=C_R+C_L$.

For the fully occupied lattice, i.e. $C=1$, the trajectories of 
the moving particle can be mapped onto the perimeters of bond 
percolation clusters$^{(4,7)}$ which was first used by 
Grassberger$^{(16)}$. A more detailed explanation about this mapping 
will be presented in the following section by relating the rotator 
model to a mirror model. Since the critical concentration $p_c$ 
for bond percolation on the square lattice is $1/2$, the critical
 concentration for the rotator model is also $1/2$, i.e. $C_{R_
c}=C_{L_c}=1/2$. Note that the right rotators are either on the 
outer side of the trajectories, in which case the trajectory is 
traversed clockwise, or on the inner side of the trajectories, 
in which case
the trajectory is traversed counter-clockwise (Fig. 3a). It is 
convenient, in our numerical simulations, to introduce for the 
analysis of the trajectories the winding angle W, which, for example
 allows a determination of whether a closed trajectory is  
traversed clockwise or counter-clockwise. For a given trajectory, 
$W$ is computed, at each step, by increasing it by $\pi/2$ if the
 particle is turned to the right, and decreasing it by $\pi/2$ 
if the particle is turned to the left. Then, when the 
trajectory closes, the winding angle will either be $2\pi$ or $-
2\pi$ which corresponds to clockwise or counter-clockwise rotation
, respectively, because trajectories cannot cross themselves.

Although we expect that, for large trajectories, the number of 
right rotators $N_R$ and the number of left rotators $N_L$ contained
 in the trajectories are on average the same, there are symmetric
 fluctuations of $N_R/N$ with respect to the mean value $C_{
R_c}=1/2$. The distribution of $N_R/N$ can be fitted to a double
 Gaussian (Fig. 4a). The symmetry is due to the fact that the 
same trajectory can be generated by replacing all right (left) 
rotators with
left (right) rotators, respectively, and reversing the particle 
velocity. However, if we just look at the clockwise traversed
trajectories, we find that the fluctuations of $N_R/N$ are no 
longer symmetric with respect to 
$1/2$ (Fig. 4b). This is not difficult to understand, because the
 right rotators are always on the outer side of the trajectories
 while the left rotators are always on the inner side of the 
trajectories. In other words, the symmetry between right rotators 
and left rotators is broken by our selection. From our numerical
 results, we found that for most clockwise closed trajectories $
N_R>N_L$, while in a few cases, $N_R<N_L$ (Fig. 4b). As we 
increase the size of the trajectories, the average of the 
distribution of $N_R$ and $N_L$ shifts toward $1/2$ with a power
 law (Fig. 5),
\begin{eqnarray}
<\frac{N_R}{N}-1/2>_c=<\frac{N_L}{N}-1/2>_{cc} \sim N^{-0.57}   
 \label{sq1}
\end{eqnarray} 

It has been have noticed before$^{(4,9)}$ that memory effects play
a dominant role in the motion of the particle on the lattice 
with fixed scatterers and in the generation of its trajectory. 
However, no direct measurement of a memory effect has been given.
 In this paper, we do so for the rotator model, by examining how
 many sites on a trajectory are
visited once $N_{1}$ (no memory effect) and how many sites on the
 same trajectory are visited twice $N_{2}$ (memory effect). A site
 on the trajectory can not be visited more than twice on the fully 
occupied lattice, since for a particle to return to the same site, it
 has to move an even number of steps. As at each step, the winding
 angle changes by $\pm\pi/2$, the change of the winding angle 
can only be either $2\pi$ or $\pi$ when the particle returns to 
the same site, (Fig. 3), so that only $N_1$ and $N_2$ are allowed.
 Thus we have the following sum rule,
\begin{eqnarray} 
N_{1}+N_{2}=N  \label{sq2}
\end{eqnarray}
If $S$ is the length of the trajectory, we also have,
\begin{eqnarray} 
N_{1}+2N_{2}=S  \label{sq3}
\end{eqnarray}
Our numerical results show that (Fig. 6),
\begin{eqnarray}
<\frac{N_{1}}{N}-\frac{1}{2}>=<\frac{1}{2}-\frac{N_{2}}{N}> \sim
 N^{-0.57}
\label{sq4}
\end{eqnarray}
Note that $N_1/N$ approaches $1/2$ from above while $N_2/N$ approaches
 $1/2$ from below. i.e. 
for infinitely large trajectories $N_1/N$ and $N_2/N$ are equal.
 This implies that memory effects are indeed very important even
 for those trajectories whose length are infinitely large. The 
asymptotic value $1/2$ for $N_1/N$ and $N_2/N$
was noticed independently by Ziff$^{(17)}$. 

Besides the memory effect exhibited by $N_1$ and $N_2$, there are
 other interesting structural properties associated with $N_1$ 
and $N_2$. For each site belonging to $N_1$, only two out of four
 adjacent bonds belong to the trajectory, while for each site 
belonging to $N_2$, all four adjacent bonds belong to the trajectory
(Fig. 3a), i.e. the site is surrounded by the trajectory (Fig. 3a). 
In other words, the $N_1$ form the ``surface'' of a trajectory,
 while the $N_2$ form the ``bulk'' of the trajectory. We 
want to point out that the ``surface'' and the ``bulk'' introduced
 here are different from the outer and inner sides we introduced
 earlier, since some sites on the outer side of a trajectory
belong to $N_2$ and some sites on the inner side of that 
trajectory belong to $N_1$. 

Although Eq.~(\ref{sq1}) and Eq.~(\ref{sq4}) exhibit
the same power law behavior, we have not obtained an explanation
 for this. From Eq.~(\ref{sq2}), Eq.~(\ref{sq3}) and Eq.~(\ref{sq4}), the 
following relation can be derived,
\begin{eqnarray}
<\frac{S}{N}-\frac{3}{2}> \sim  N^{-0.57} \label{sq5}
\end{eqnarray}
indicating that $S$ and $N$ are asymptotically proportional to 
each other. Therefore, replacing $S^{-0.57}$ by $N^{-0.57}$ in all
 scaling equations is justified, as long as the asymptotic 
behavior is considered. 

\subsection{The partially occupied lattice}
For a partially occupied lattice, the trajectories can cross 
themselves (Fig. 2a). The outer side and inner side of the closed 
orbits cannot be well defined anymore, so that 
a direct mapping of the trajectories onto the perimeters of bond
 percolation clusters is not possible$^{(7,18)}$. Nevertheless 
earlier numerical studies$^{(7)}$ have still shown the existence 
of two critical lines symmetric with respect to the line $C_R=C_
L$ (Fig. 7) and the critical exponents,
$\tau$ and $d_f$ have the same values as at $C=1$.  However, the 
existence of these critical lines below $C = 0.56$ could not be established,
due to the prohibitively long numerical calculations needed. We will give
more details about some special features of the critical lines in
 the following paper. 

One important consequence of the inequality
 of $C_R$ and $C_L$ at criticality for $C<1$ is that the winding
 angles are asymptotically proportional to the number of sites
 or to the number of time steps contained in the closed trajectories.
 As an example, we consider to the critical point at concentration
 $C=0.90$, namely, $C_R=0.477$, $C_L=0.423$. To our surprise,
 our numerical results exhibit that the deviation of the 
winding angle from its mean value $(C_R-C_L)$ decays not with a 
power law but with a stretched exponential law (Fig. 8),
\begin{eqnarray}
<|\frac{W(N)}{N}-(C_R-C_L)|> \sim 2^{(-2.8 \, N^{0.18})} 
\label{dsqw}
\end{eqnarray} 
For $N_R/N$, $N_L/N$ and $N_E/N$, our numerical calculations yield
 the following scaling relation (Fig. 9),  
\begin{eqnarray}
<|\frac{N_R}{N}-C_R|>
\sim <|\frac{N_L}{N}-C_L|> 
\sim <|\frac{N_E}{N}-C_E|> 
\sim N^{-0.50}   \label{dsq0}
\end{eqnarray}
Note that the value of the exponent $-0.50$ differs from that for
 the fully occupied lattice $-0.57$, indicating that self-crossing
 occurs randomly as the particle generates its trajectory.
  
The memory effects on the partially occupied lattice are more 
complicated than on the fully occupied lattice. For example, the 
number of times that a site can be visited by the moving particle
 can range from one to four. We denote the number of these different
 types of sites by $N_1$, $N_2$, $N_3$ and $N_4$,  respectively.
 An interesting feature for $C<1$ is that ``reflectors'' can
 be formed after a number of time steps on a trajectory (a 
``reflector'' is formed when the moving particle travels through the 
same bond as it has traveled on before, but in the opposite direction
, Fig. 10. We remark here that these ``reflectors'', are 
responsible for the absence of diffusion for the flipping rotator 
model as proved by Bunimovich and Troubetzkoy$^{(19)}$ and make 
$N_3$ and $N_4$ non vanishing. Since each site belongs to only one
 of the four types, $N_1$, $N_2$, $N_3$ or $N_4$, we have the 
following two sum rules,
\begin{eqnarray}
N_1+N_2+N_3+N_4=N  \label{dsq1}
\end{eqnarray}
and
\begin{eqnarray}
N_1+2N_2+3N_3+4N_4=S  \label{dsq2}
\end{eqnarray}
These two relations are quite useful for numerical simulations, 
since for known $N_1$ $N_2$, $N$ and $S$, one can find $N_3$ and
 $N_4$, saving therefore a large amount memory and computational
 time. If the asymptotic values for $N_1/N$, $N_2/N$, $N_3/N$  
and $N_4/N$ are represented by $K_1$, $K_2$, $K_3$ and $K_4$, 
respectively, our numerical simulations show the following power law
 behavior (Fig. 11),
\begin{eqnarray}
<\frac{N_1}{N}-K_1> \sim  
<\frac{N_2}{N}-K_2> \sim   
<K_3-\frac{N_3}{N}> \sim     
<K_4-\frac{N_4}{N}> \sim N^{-0.39} \label{dsq3}
\end{eqnarray}
The sum of $K_1$, $K_2$, $K_3$ and $K_4$ is equal to one as 
required by Eq.~(\ref{dsq1}). We found that $K_1$, $K_2$, $K_3$ and 
$K_4$ are functions of the concentration $C$ and all converge to
 values near $1/4$, (Fig. 12). The exponent $-0.39$ in the above
 equation is significantly different from the corresponding one 
for the fully occupied lattice $-0.57$, although $\tau$ and $d_f$
 are still the same as found by Cohen and Wang$^{(7)}$. The same
 scaling behavior is found at $C=0.80$, suggesting that the critical
 behavior is universal along the critical line. 

In the same way as was done for the fully occupied lattice, we can
 derive the scaling behavior for $S/N$ from Eq.~(\ref{dsq1}), 
(\ref{dsq2}) and (\ref{dsq3}),
\begin{eqnarray}
<\frac{S}{N}- 2.3670>\sim N^{-0.39} \label{dsq4}
\end{eqnarray}
We note that the sites that belong to $N_2$ can be separated into
 two classes: those that only have three adjacent bonds traversed
 by the particle and those that have four adjacent bonds traversed
 by the particle. The former together with $N_1$ form the
``surface'' of the trajectories while the latter together with $N_
3$ and $N_4$ form the ``bulk'' of the trajectories.

\section{Critical behavior of the mirror model on the square 
lattice}
\subsection{The fully occupied lattice}
The mirror model was proposed by Ruijgrok and Cohen sometime ago
$^{(1)}$, where the scatterers consist of right mirrors (tilted 
to the right by $\pi/4$, Fig. 1c) and left mirrors (tilted to the
 left by $\pi/4$, Fig. 1d), which reflect the particle upon 
collision, like a photon is reflected from a mirror. $C_R$ and $C_L
$ are now the concentrations of right mirrors and left mirrors, 
respectively, and $C=C_L+C_R$. It has been shown before$^{(4,7,2
0)}$ that $C=1$ is a critical line for the mirror model. 
Only for $C_R=C_L=1/2$, can the mirror model and the rotator model
 be mapped onto each other. That is, by properly replacing right
 mirrors and left mirrors 
with either left rotators or right rotators, we can produce the 
same 
trajectory with the same probability. For $C_R\neq C_L$, the 
trajectories can still be mapped onto each other, but the probability
 to generate the same trajectory by mirrors is 
no more the same as by rotators, so that these two models cannot
 be mapped onto each other anymore. However, we will show later 
in this section, that the mirror model can still be related to a
 rotator model, even when $C_R \neq C_L$.

The mirror model can, for all $C=1$, still be mapped  onto an
isotropic bond percolation problem at criticality, (except in the
 two limits: $C_R=1$, $C_L=0$, and $C_R=0$, $C_L=1$, where the 
moving particle simply zig-zags to infinity). Therefore the trajectory
length distribution, characterized by $\tau=15/7$ and the  fractal 
dimension $d_f=7/4$ are still the same as before. However 
$\sigma$ cannot be defined anymore for the mirror model at $C=1$, 
because we are on a critical line rather 
than at a critical point, as was the case for the rotator model.

Unlike in the rotator model, in the mirror model right mirrors 
can either be on the outer or on the inner sides of a trajectory,
 therefore the method of separating trajectories into clockwise 
and counter-clockwise ones will not break the symmetry between 
the number of right mirrors $M_R$ and the number of left 
mirrors $M_L$ on the trajectories. Indeed, our numerical results show 
that the distribution of $M_R/N$ ($N=M_R+M_L$) is a Gaussian centered
 at $C_R$ rather than a double Gaussian as we found in the 
rotator model, Fig. 13. The following power law describes how this
 distribution narrows as $N$ increases (Figs 13,14), 
\begin{eqnarray}
<|\frac{M_R}{N}-C_R|> = <|\frac{M_L}{N}-C_L|> \sim N^{-0.50}
\end{eqnarray}
The exponent $0.50$ is not surprising here. The key point is that $C=1$ is a 
critical line, so that for a large trajectory, we cannot tell at
 which concentration it is generated.
In other words, the number $M_R/N$ can fluctuate freely about its mean 
value $C_R$. According to the central limit theorem, we get then
 the observed result.

Memory effects can be studied in the same way as for the rotator
 model, because each site on the trajectories can be only visited either
 once or twice, the same as for the fully occupied rotator model.
Our numerical results show that the scaling behavior of $N_1/N$ 
and $N_2/N$ is the same as that for the rotator model, except that
 the asymptotic value now depends on the concentration $C_{R}$, (Fig. 15),
\begin{eqnarray}
<\frac{N_1}{N}-a(C_R)> = <b(C_R)-\frac{N_2}{N}>\sim N^{-0.57}  \
\label{ch3eq1}
\end{eqnarray}
From $N_1+N_2=N$, we have $a(C_R)+b(C_R)=1$. 
For $C_R>1/2$, our simulations show that, $a(C_R)$ is monotonically
decreasing with $C_R$,
while for $C_R<1/2$, $a(C_R)$ is monotonically increasing with $
C_R$. At $C_R=C_L=1/2$, we recover the result of the rotator model,
i.e. $a(1/2)=b(1/2)=1/2$.
In fact, $a(C_R)$ can be fitted very well to a parabola (Fig. 16
) except at $C_R=0$ and $C_R=1$, where $N_1/N=1$ and $N_2/N=0$ 
indicating a discontinuous transition at these two limits.
 
The scaling behavior for $S/N$ can be derived from the relation 
$N_1+N_2=N$, $N_1+2N_2=S$, and Eq.~(\ref{ch3eq1}),
\begin{eqnarray}
< \frac{S}{N}-[1+b(C_R)] >\sim N^{-0.57}
\end{eqnarray}
This scaling behavior is the same as that found for the fully 
occupied rotator model, Eq.~(\ref{sq5}), except that the asymptotic
 value, is now a function of $C_R$, rather than a single value 
$3/2$, as for the rotator model. 

As we mentioned before, each mirror on a given trajectory can be
 replaced by an 
appropriately chosen rotator to reproduce the same trajectory. 
After this operation, one can see that all right rotators are 
either on the outer side or on the inner side of the trajectory 
depending on the velocity direction of the particle, the same feature
 we found for the rotator model on a fully occupied lattice. We
 will refer to this rotator model generated from the mirror model
 as a quasi-rotator model. It is natural to 
study, in the same way as for the rotator model, the scaling 
behavior of the number of right rotators $N_{R}$ and 
left rotators $N_L$ on the trajectories for the quasi-rotator
model. Our 
numerical results indicate the same power law behavior is obtain
ed as in the rotator model, even when $C_R \neq C_L$ where the 
rotator model and the quasi-rotator (or mirror model) cannot be 
mapped onto each other, (Fig. 17).
\begin{eqnarray}
<\frac{N_R}{N}-\frac{1}{2}>_c = <\frac{N_L}{N}-\frac{1}{2}>_{cc}
 \sim N^{-0.57}
\end{eqnarray}   
The asymptotic values for $N_R/N$ and $N_L/N$ are equal and the 
same as for the rotator model on the fully occupied lattice. This
 is not surprising, because $N_R$ is associated with the outer 
side of a trajectory, while $N_L$ is associated with the inner side
 of a trajectory (Fig. 3b). Although the mirror model for all
 $C=1$ corresponds to a quasi-rotator model, it is not clear how
 the quasi-rotator model can be generated directly, other than 
deriving it from a mirror model.  

\subsection{The partially occupied lattice}
Here we only give a very brief discussion of the mirror model on
 a partially occupied square lattice.  Again, this model can not
 be mapped onto a percolation model due to the fact that 
trajectories can cross themselves (Fig 2b). At $C_R=C_L=C_E=1/3$, the 
mirror model can be mapped onto a growing self-avoiding walk trail
 introduced by Lyklema$^{(21)}$ some time ago. Previous studies
 have shown that the distribution of the size of closed trajectories 
and the fractal dimension of large trajectories are drastically
 changed as soon as $C<1$: $\tau=1$ and $d_f=2$, with logarithmic
corrections$^{(4,7,22,23)}$, indicating a first order phase
 transition.
It has also  been shown by numerical simulations that the moving
 particle exhibits super-diffusive behavior$^{(7,23)}$, i.e. the
 diffusion constant increases with time logarithmically, suggesting
 that the hyperscaling relation $\tau-1=2/d_f$ needs logarithmic
 corrections. It is not difficult to show that no ``reflector''
 can be formed in this model and consequently, that each site 
on a trajectory can only belong to either $N_1$ or $N_2$.   
\section{Critical behavior of the rotator and the mirror model on
 the triangular lattice}
\subsection{The fully occupied lattice}
In order to investigate the universality of the trajectory scaling
 behavior found on the square lattice, we also studied the 
triangular lattice. Similar to the rotator and mirror model on the 
square lattice, fixed rotators or fixed mirrors are randomly placed
 on the sites of the triangular lattice. The particle moves in
 six directions along the bonds of the lattice and its velocity
 is rotated by scatterers over an angle of $\pm 2\pi/3$ upon each
 collision, (Fig. 18). Since both the mirror model and the rotator
 model are discussed in this section, in order to avoid confusion
, we define $C_R^{rotator}$ and $C_L^{rotator}$ as the 
concentration for right rotators and left rotators, respectively, 
while we define $C_R^{mirror}$ and $C_L^{mirror}$ as the concentration
for right mirrors and left mirrors, respectively. Similarly 
as before, $C$ represents the total concentration of the scatterers.

For $C=1$, earlier papers have shown$^{(3,7)}$ that the mirror 
model and the rotator model can be mapped into each other by replacing
 all the right (left) mirrors on the trajectories by either right (left)
rotators or left (right) rotators, respectively, provided $C_R^{rotator}=C
_R^{mirror}$ and $C_L^{rotator}=C_L^{mirror}$. Note that this is
 a global transformation rather than a local transformation, as 
was the case with the mapping of the rotator model onto the mirror
 model on the square lattice for $C_R = C_L = 1/2$ (Fig. 19).
In other words, the mirror model and the rotator model are equivalent
. Again it can be shown that the trajectories cannot cross 
themselves and that all right rotators are either on the outer 
side or the inner side of the trajectories, depending on the 
direction in which the particle trajectory closes.

Both the mirror model and the rotator 
model can be mapped onto a site percolation rather than a bond 
percolation problem, as was the case for the rotator and the mirror
 models on the square lattice. For site percolation on the 
triangular lattice, the critical probability $p_c$ to occupied a 
lattice site is $1/2$, so that the critical point occurs for $C_L^
{mirror}=C_R^{mirror}=C_L^{rotator}=
C_R^{rotator}=1/2$. Here, we will only consider the rotator model,
 since the mirror model can be done in the same fashion.
The number of right rotators $N_R$ and left rotators $N_L$ contained
 in the closed trajectories, exhibit the same power law 
behavior as on the square lattice, (Fig. 20),
\begin{eqnarray}
<\frac{N_R}{N}-\frac{1}{2}>_c=<\frac{1}{2}-\frac{N_L}{N}>_{cc} 
\sim N^{-0.57}
\label{Tr1}
\end{eqnarray}
where $N=N_R+N_L$. The average is taken over all trajectories 
which contain $N$ sites. 

For the triangular lattice, the maximum number of times a site on a
trajectory can be visited by the moving particle is three.
This can be shown by considering again the winding angle $W$. Upon
 each collision $W$ will either increase or decrease by $2\pi/
3$. Therefore, when the particle returns to the same site, the 
winding angle will be a multiple of $2\pi/3$. However, there are 
only three allowed angles of the form $2n\pi/3$ on the triangular
 lattice, where $n$ can be $0$, $1$ or $2$, so that the maximum
 number of times a site be visited by the moving particle is indeed
 three. 
We define, $N_1$, $N_2$ and $N_3$ as the number of sites visited
 by the particle once, twice, three times and four times, 
respectively, while $S$ is the length of the trajectory. These quantities
 satisfy the following two sum rules,
\begin{eqnarray}
N_1+N_2+N_3=N    \label{Tr2}
\end{eqnarray}
and
\begin{eqnarray}
N_1+2N_2+3N_3=S \label{Tr3}
\end{eqnarray} 
The scaling behavior of $N_1/N$, $N_2/N$ and $N_3/N$ is very 
similar to that found on the square lattice (Fig. 21),             
                             
\begin{eqnarray}
<\frac{N_1}{N}-K_1> \sim 
<K_2-\frac{N_2}{N}> \sim 
<K_3-\frac{N_3}{N}>\sim N^{-0.57}
\label{Tr4}
\end{eqnarray}
where $K_1=0.3197$, $K_2=0.4052$, $K_3=0.2751$. The sum of $K_1$
, $K_2$ and $K_3$ is one as required by Eq.~(\ref{Tr2}). Although
 all values of $K_i$ $(i=1,2,3)$ are close to $1/3$, they are not
 equal. 
Since each site has six adjacent bonds, the sites that belong to
 $N_1$ and $N_2$ form the ``surface'' of the trajectories, while
 the other sites that belong to $N_3$ form the ``bulk'' of the 
trajectories. From Eq.~(\ref{Tr4}), we see that the ``surface'' is
 much bigger than the ``bulk''. 
The asymptotic scaling behavior for $S/N$ follows from Eq.~(\ref
{Tr2}), Eq.~(\ref{Tr3}) and Eq.~(\ref{Tr4}), to be,
\begin{eqnarray}
<S/N-1.9554> \sim N^{-0.57}
\end{eqnarray}

\subsection{The partially occupied lattice}
For the partially occupied lattice, i.e. $C<1$, the rotator model
 and the mirror model can still be mapped into each other by a 
global transformation. That is, for a given trajectory,
all left (right) mirrors on the trajectories can be replaced by
 either left (right) rotators or right (left) rotators respectively
 while empty sites remain empty, (Fig. 22).
However, neither of the two models can be mapped onto a percolation
problem, because the trajectories can cross themselves via the
empty sites and consequently, the inner side and the outer side
 of large trajectories cannot be defined properly anymore, as was
 also the case for the square lattice partially occupied by 
scatterers.
Nevertheless it was found numerically by
Cohen and Wang$^{(7)}$ that for $C<1$, there exists a linear critical
 line $C_R=C_L=C/2$ (Fig.~23) and the cluster perimeter 
distribution exponent $\tau=15/7$ and the fractal dimension $d_f=7/
4$
of the trajectories along this critical line belong to the same 
universality class as that of site percolation. However, the 
existence of this simple critical line at low concentrations could
 not be established due to the prohibitively long numerical 
calculations needed. The lowest concentration studied in our simulations
 was $C_R=C_L=C/2=0.225$.

Since the trajectories can cross themselves, the winding angle $
W$ for closed trajectories is no more restricted to $2\pi$ or $-
2\pi$, so that it is natural to study the scaling behavior of $W
$. Our numerical results (Fig. 24) for $C_R=C_L=C/2=0.45$ show the
 following relation,
\begin{eqnarray}
<|\frac{W(N)}{N}|> \sim N^{-0.50}
\end{eqnarray} 
in contrast to the stretched exponential behavior for the partially
 occupied rotator on the square lattice, Eq.~(\ref{dsqw}). The
 asymptotic value for $W(N)/N$ is zero, because the moving particle
 has equal probability to turn to its right or to its left.

For $N_R$, $N_L$ and $N_E$, our numerical simulations yield a 
different scaling exponent from that for the fully occupied lattice
, (Fig. 25),
\begin{eqnarray}
<|\frac{N_R}{N}-C_R|> \sim <|\frac{N_L}{N}-C_L|> \sim <|\frac{N_
E}{N}-C_E|> \sim N^{-0.50}
\end{eqnarray}
This scaling behavior is the same as that of the partially occupied
 square lattice, Eq.~(\ref{dsq0}). The exponent $-0.50$ can be 
understood in the same way as for the mirror model for $C=1$.

Using the same kind of argument, as we used above for the fully 
occupied 
lattice, i.e. considering the change of the winding angle,  we can
 show that the maximum number that a site on a trajectory
can be visited by the moving particle is three. Unlike the rotator
 model on the  partially occupied  square lattice, 
Eq.~(\ref{dsq3}), we find,
that $N_1/N$, $N_2/N$ and $N_3/N$ now
follow the same scaling behavior along the critical line as 
 the rotator model on the fully occupied lattice, (Fig. 26),
\begin{eqnarray}
<\frac{N_1}{N}-K_1> \sim <K_2-\frac{N_2}{N}> \sim <K_3-\frac{N_3
}{N}> \sim N^{-0.57}
\end{eqnarray}
where $K_1$, $K_2$, $K_3$ are functions of the concentration, 
which sum to 1, while the exponent $-0.57$ is independent of the concentration. 
Our numerical simulations also show that $K_1$ is a not a monotonic
 function of $C$ while $K_2$ is an increasing function of $C$ 
and $K_3$ is a decreasing function of $C$. However, all  $K_i$ ($
i=1,2,3)$ appear to converge to three different constants as $C$
 decreases, (Fig. 27).
\section{Conclusion}
In this paper, we have given, a more detailed
 numerical analysis than before of the nature of the trajectories
 generated in a 
Lorentz lattice gas model on both the square lattice and the 
triangular lattice at criticality.
Our study has shown that the ``structural'' properties of these 
trajectories are highly non-trivial and yield some new
universal scaling behavior probably related to the scaling function
 exponent $\sigma$, in addition to that associated with the 
critical exponents $\tau$
and $d_{f}$ found before. We hope that this study can lead to a 
more complete understanding of the complicated trajectories 
generated by a particle moving through randomly distributed scatterers
 on a lattice. The results are summerized in table I, table II
 and table III.

Our study only answered a few questions, but raised many more 
questions, as yet unanswered, among which we note the following. 

\noindent 1. For $C=1$, why are the asymptotic values of $N_R$ 
and $N_L$ as well as of $N_1$ and $N_2$ on the fully occupied square
 lattice equal to $N/2$, i.e. why is $K_1=K_2=1/2$? What determines
 the asymptotic values for $K_1$, $K_2$ and $K_3$ on the fully
 occupied triangular lattice? 

\noindent 2. Our numerical simulations indicate that the $K_i$ (
$i$ can be as large as $4$ for the rotator model on the partially
 occupied square lattice and $3$ for the triangular lattice) 
approach constants, as $C$ is decreased along the critical line; 
in other words, the structure of the large trajectories hardly 
changes as the concentration of the scatterers is reduced, but why?

\noindent 3. Although we have given an argument why the exponents
 describing the fluctuations of $N_R /N$ along 
the critical line, $C=1$ for the mirror model, are all equal to 
$0.50$, the more fundamental exponent $0.57$ still remains elusive.
We conjecture that
the exact value is $1-\sigma=4/7$, but it is not clear how to 
obtain this from theory.

\noindent 4. Why does the winding angle $W$ on the rotator model
 on the partially occupied square lattice exhibit a stretched 
exponential law behavior, while it exhibits a simple power law 
behavior on the partially occupied triangular lattice?     

\noindent 5. We find, that the $N_i/N$ approach their asymptotic
 values $K_i$ with a power law $N^{-0.57}$ in all cases, except 
for the rotator model on the partially occupied square lattice 
where the exponent has a significantly different value, $-0.39$. 
What is the origin of this difference? In the next paper, dealing
 with scaling behavior near criticality, we will show that 
$\sigma$ has the same value $3/7$ in all cases except again for the 
partially occupied rotator model on the square lattice.  This 
may indicate that the scaling behavior of the $N_i/N$ and the exponent
 $\sigma$ are related to each other.

\noindent 6.  How can one understand the difference between the 
phase diagram of the partially occupied rotator model on the square
 lattice and that on the triangular lattice? Do the ``reflectors''
 play an important role here?

\noindent 7. How are the lattice sites visited by a moving particle,
 once, twice, etc. correlated to each other on the lattice, 
with respect to both space and time? \\ 

\noindent{\bf {Acknowledgment} }

The authors are grateful to J. Machta and F. Wang for useful 
discussions, and R. Ziff for a number of helpful comments on the 
manuscript. This work is supported under grant DE-FG02-88ER13847 
of the Department of Energy.
\newpage
\noindent{\bf References}

\noindent 1. Th. W. Ruijgrok and E. G. D. Cohen, Phys. Lett. A 1
{\bf{33}}, 415 (1988).

\noindent 2. X. P. Kong and E. G. D. Cohen, Phys. Rev. B {\bf{40}}, 483
8 (1989).

\noindent 3. X. P. Kong and E. G. D. Cohen, J. Stat. Phys. {\bf{62}}, 7
37 (1991).

\noindent 4. R. M. Ziff, X. P. Kong and E. G. D. Cohen, Phys. Rev.
 B {\bf{44}}, 2410 (1991).

\noindent 5. E. G. D. Cohen, New types of diffusion in lattice gas
 cellular automata, in {\em Microscopic Simulations of Complex 
Hydrodynamic Phenomena}, M. Mar\'{e}chal and B. L. Holian, eds. 
(Plenum Press, New York, 1992), p. 137.

\noindent 6. H. F. Meng and E. G. D. Cohen, Phys. Rev. E {\bf{50}}, 248
2 (1994).

\noindent 7. E. G. D. Cohen and F. Wang, J. Stat. Phys. {\bf{81}}, 445 
(1995).

\noindent 8. F. Wang and E. G. D. Cohen, J. Stat. Phys. {\bf{81}}, 467 
(1995).

\noindent 9. E. G. D. Cohen and F. Wang, Physica A {\bf{219}}, 56 (1995).

\noindent 10. E. G. D. Cohen and F. Wang, Fields Institute Communications,
 6, 43 (1996).

\noindent 11. B. Sapoval, M. Rosso and J. F. Gouyet, J. Phys. Lett.
 (Paris) {\bf{46}}, L149 (1985).

\noindent 12. R. M. Ziff, Phys. Rev. Lett. {\bf{56}}, 545 (1986).

\noindent 13. H. Saleur and B. Duplantier, Phys. Rev. Lett. {\bf{58}}, 
2325 (1987).

\noindent 14. R. M. Ziff, P. T. Cummings, and G. Stell, J. Phys.
 A {\bf{17}}, 3009 (1984).

\noindent 15. J. Gunn and M. Ortun\~{o}, J. Phys. A {\bf{18}}, 1035 (1985).

\noindent 16. P. Grassberger, J. Phys. A {\bf{19}}, 2675 (1986).
 
\noindent 17. R. M. Ziff. Private Communication.

\noindent 18. M. Ortun\~{o}, J. Ruiz and M. F. Gunn, J. Stat. 
Phys. {\bf{65}}, 453 (1991).

\noindent 19. L. A. Bunimovich and S. E. Troubetzkoy, J. Stat.
Phys. {\bf{74}}, 1 (1994).

\noindent 20. L. A. Bunimovich and S. E. Troubetzkoy, J. Stat. 
Phys. {\bf{67}}, 289 (1992).

\noindent 21. J. Lyklema, J. Phys. A {\bf{18}}, L617 (1985).

\noindent 22. R. M. Bradley, Phys. Rev. B {\bf{41}}, 914 (1989).

\noindent 23. A. L. Owczarek and T. Prellberg, J. Stat. Phys. {\bf{79}},
 951 (1995).

\newpage
\noindent{\bf Figure Captions}

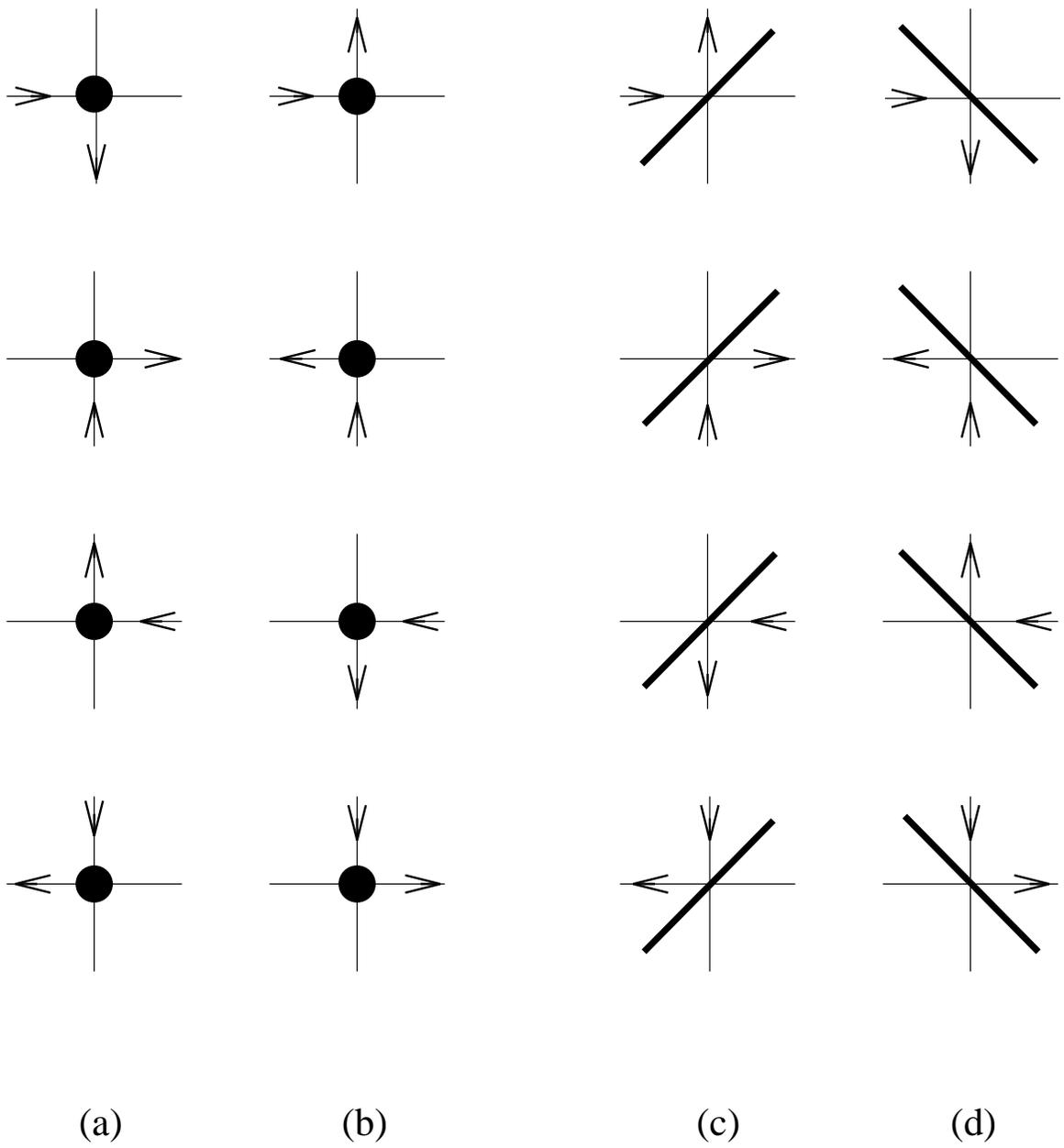
\noindent Fig. 1. Scattering rules for the rotator model and the
 mirror model on the square lattice (a) right rotators; (b) left
 rotators; (c) right mirrors; (d) left mirrors.

\noindent Fig. 2. (a) a typical closed trajectory generated by a
 moving particle on the square lattice partially occupied by 
rotators. Here $W=6\pi$, $N_R=10$, $N_L=2$, $N_E=7$, $N_1=18$, $N_2
=1$, $N_3=0$, $N_4=0$, $N=19$ and $S=20$. (b) The same trajectory
 generated by a moving particle on the square lattice partially
 occupied by mirrors. Here $M_R=8$ and $M_L=4$.

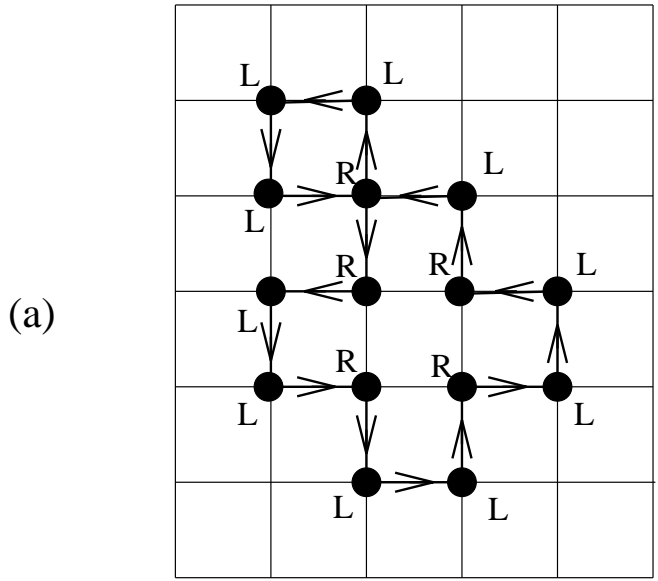
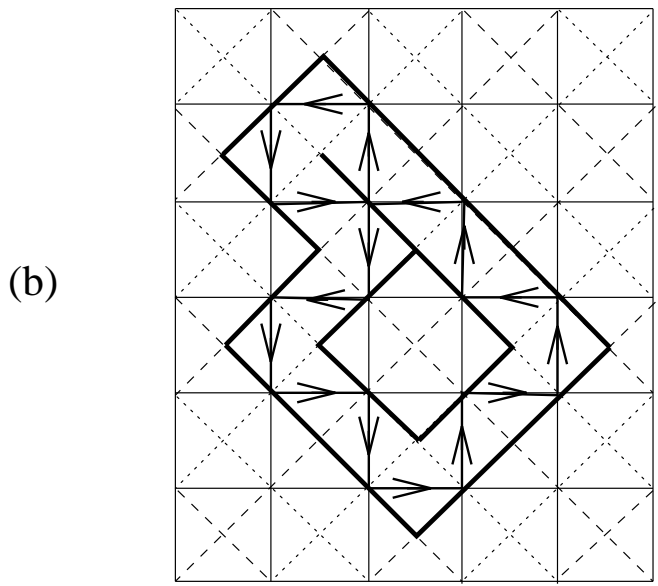
\noindent Fig. 3. (a)  A typical counter-clockwise closed trajectory
 on the square lattice. Note that all left rotators and all 
right rotators are on the outer side and inner side of the trajectory,
 respectively. Here $W=-2\pi$,  $N_R=5$, $N_L=10$, $N_1=14
$, $N_2=1$, $N=15$ and $S=16$. (b) The same trajectory generated
 by mirrors on the square lattice. Mirrors (fat solid lines) on 
the outer side of the trajectory form a bond percolation cluster
 perimeter on one of the two sublattices (dashed lines), while 
mirrors on the inner side of the trajectory form a bond percolation
 cluster perimeter on the other sublattice (dotted lines). Here
 $M_R=6$ and $M_L=9$. 

\noindent Fig. 4. The probability density $P(N_R/N)$ vs $N_R/N$ 
for the rotator model. The data were obtained from trajectories 
consisting of $1300$ to $1500$ sites ($\diamond$) and $2000$ to 
$2300$ sites ($+$). (a) both clockwise and counter-clockwise 
trajectories are included in the data. The solid and the dashed 
curves are described by the double Gaussians; $23.7 \, e^{-7058(N_R
/N-0.5182)^2}+23.7 \, e^{-7058(N_R/N+0.5182)^2}$ and $ 28.0 \, e
^{-9856(N_R/N-0.5145)^2}+ 28.0 \, e^{-9856(N_R/N+0.5145)^2}$, 
respectively. (b) only the clockwise trajectories are included in 
the data. The solid and the dashed curves are described by the 
Gaussians, $23.7 \, e^{-7058(N_R/N-0.5182)^2}$ and $ 28.0 \, e^{-
9856(N_R/N-0.5145)^2}$, respectively. For increasing values of $
N$ the curves become taller and narrower.

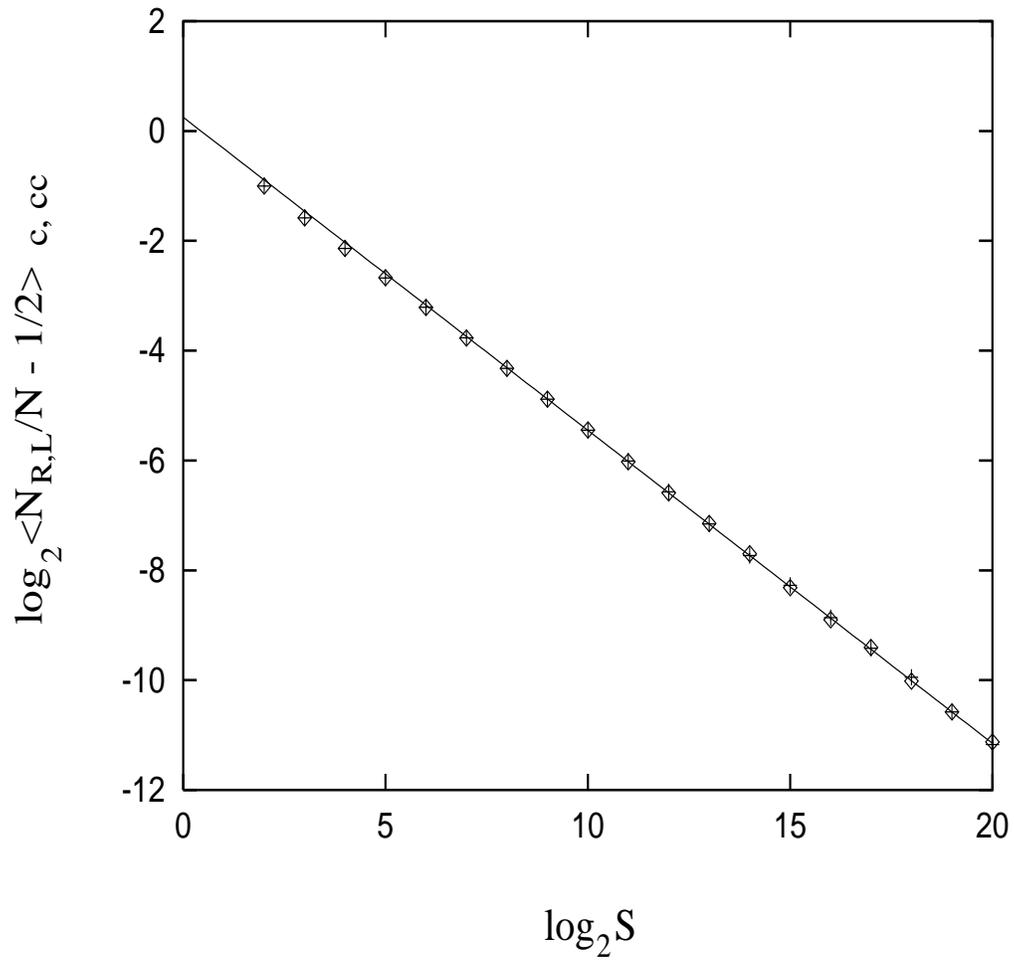
\noindent Fig. 5. Scaling behavior of $N_R$ and $N_L$ for clockwise
 and counter-clockwise closed trajectories, respectively, on 
the square lattice fully occupied by rotators, $C_R=C_L=1/2$:  
$\log_{2}< \! N_R/N-1/2 \! >_c$ vs $\log_{2}S$ ($\diamond$), 
$\log_{2}< \! N_L/N-1/2 \! >_{cc}$ vs $\log_{2}S$ ($+$). The slope of
 the lines through both sets of points is $-0.57$.

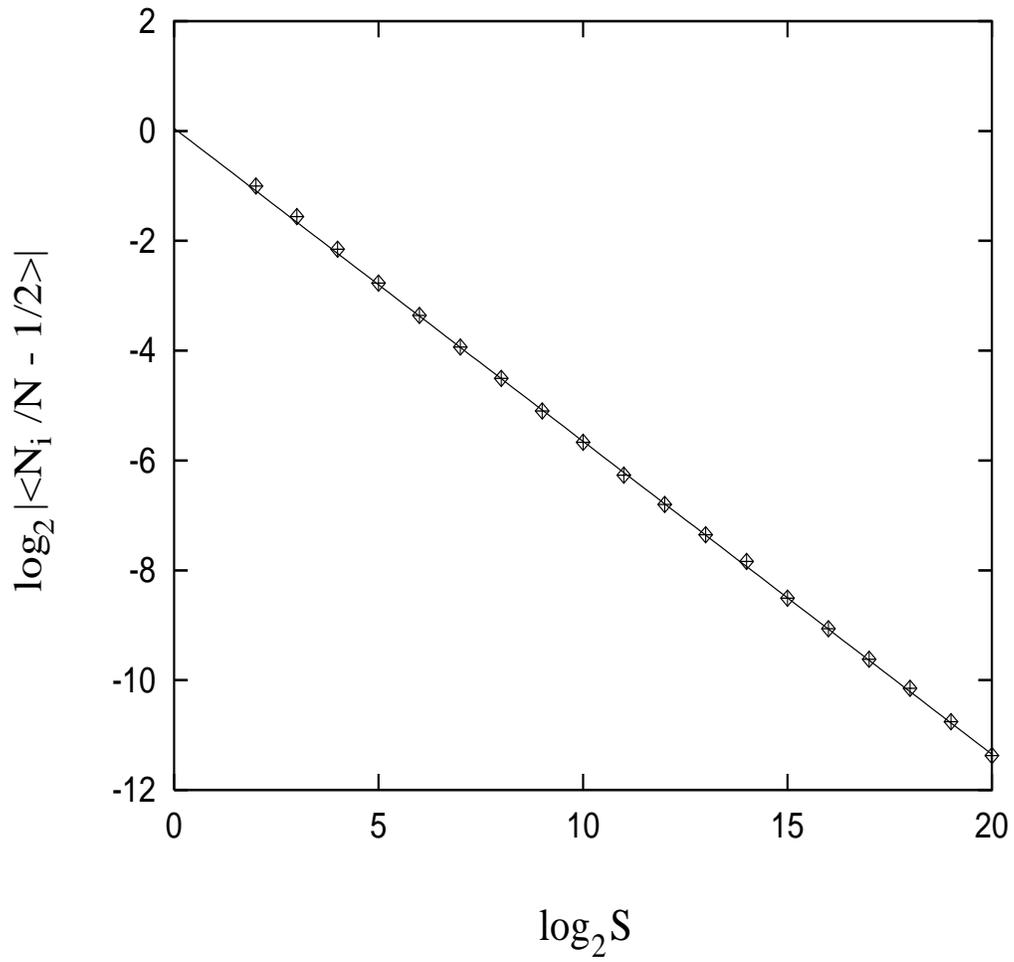
\noindent Fig. 6. Scaling behavior of $N_1$ and $N_2$ for trajectories
 on the square lattice fully occupied by rotators, $C_R=C_
L=1/2$:  $\log_{2}< \! N_1/N-1/2 \! >$ vs $\log_{2}S$ ($\diamond
$). Also plotted is  $\log_{2}< \! 1/2-N_2/N \! >$ vs $\log_{2}S
$ ($+$), which have the same values. The slope of the lines through
 both sets of points is $-0.57$.

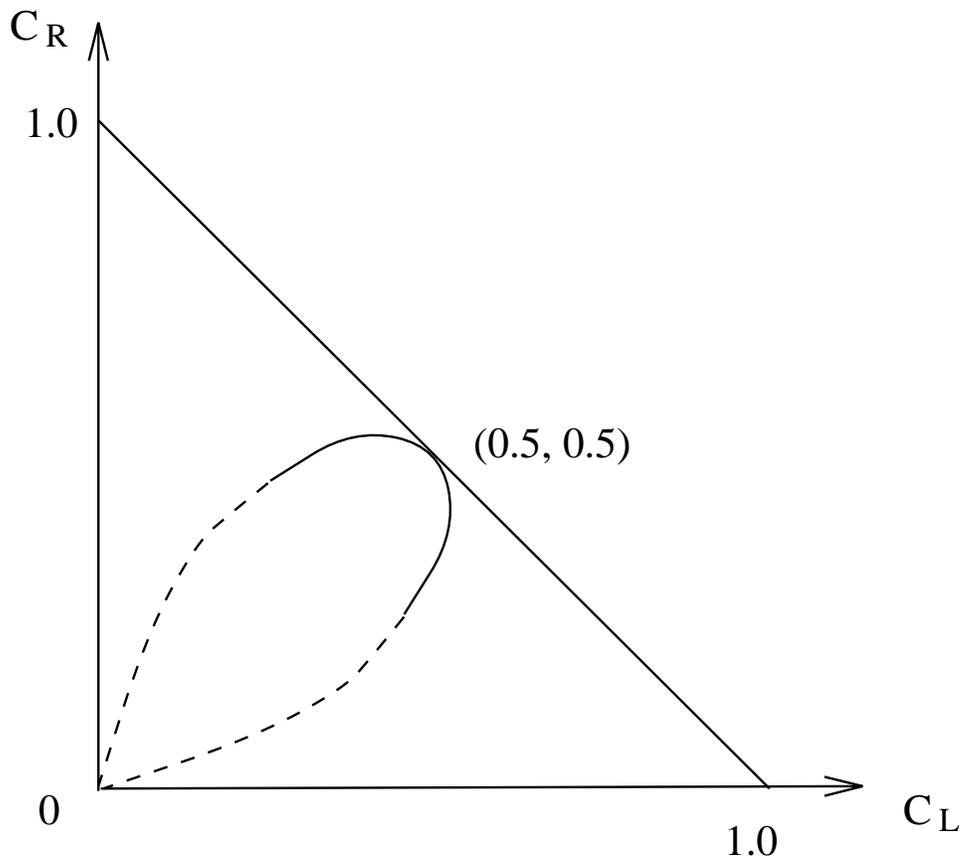
\noindent Fig. 7. Schematic phase diagram for the rotator model 
on the square lattice. The inner solid line extends to $C=0.65$ 
and is tangent to the line $C=1$, the dashed line is conjectured.

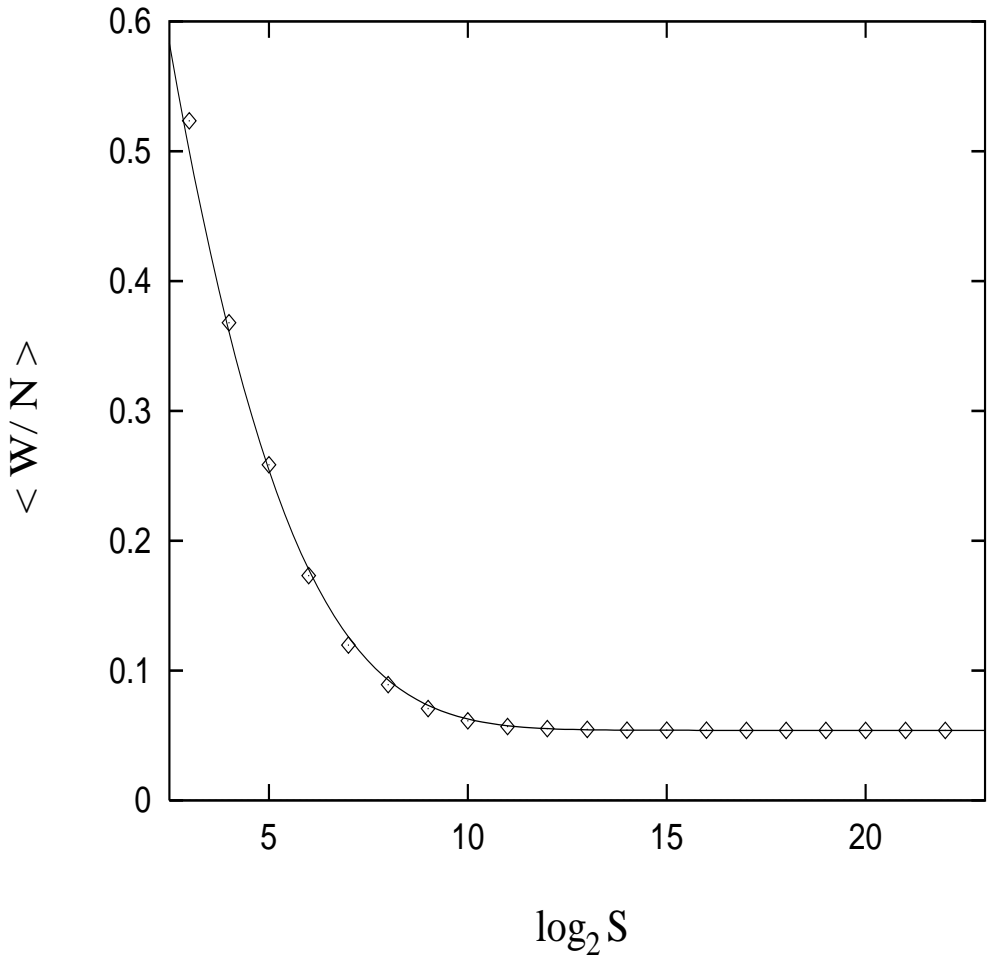
\noindent Fig. 8. Scaling behavior of the winding $W$ for trajectories
 on the square lattice partially occupied by rotators, for
 a typical concentration $C=0.9$ with $C_R=0.477$, $C_L=0.423$. 
$<\! |W/N-(C_R-C_L)| \! >$ vs $\log_{2}S$ ($\diamond$). The solid
 curve is described by $(C_R-C_L)+7.5\times 2^{(-2.80 \, S^{0.1
8})}$.

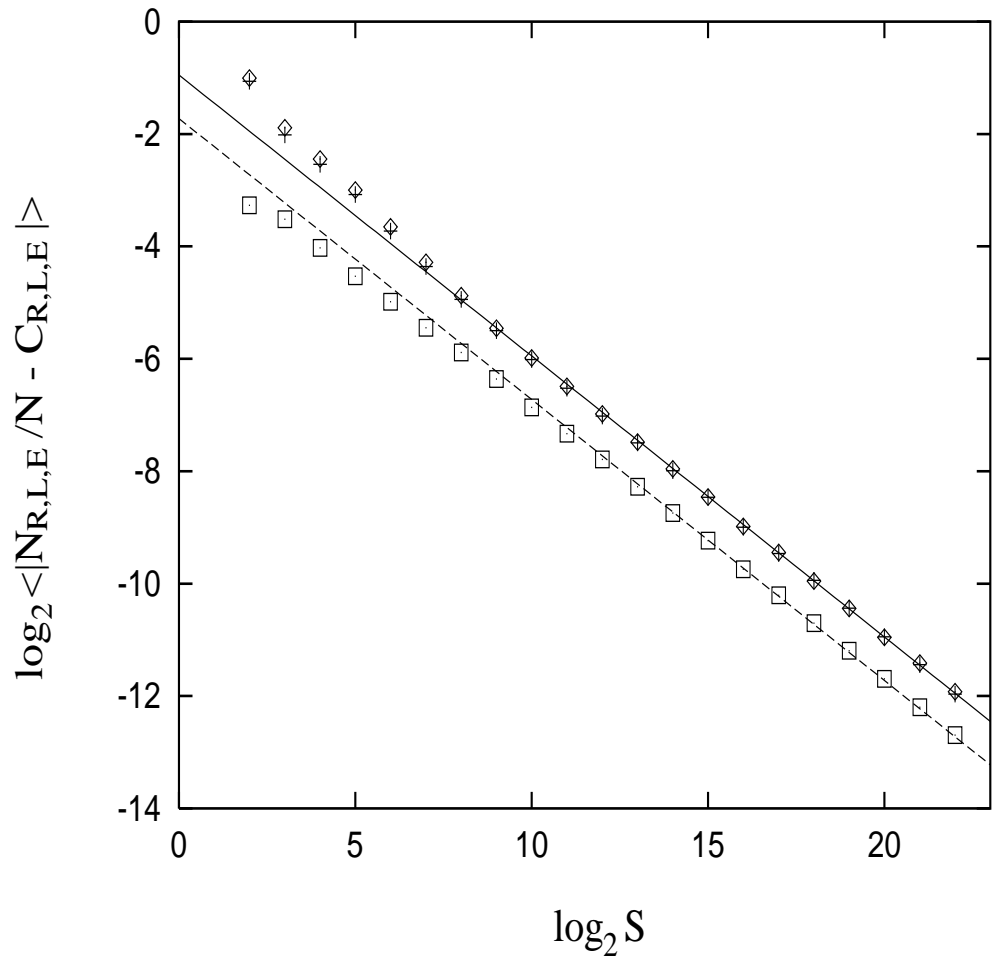
\noindent Fig. 9. Scaling behavior of $N_R$, $N_L$ and $N_E$ for
 trajectories on the square lattice partially occupied by rotators,
 for $C=0.9$ with $C_R=0.477$, $C_L=0.423$, $C_E=0.1$:  $\log
_{2}< \! |N_R/N-C_R| \! >$ vs $\log_{2}S$ ($\diamond$), $\log_{2
}< \! |N_L/N-C_L| \! >$ vs $\log_{2}S$ ($+$), $\log_{2}< \! |N_E
/N-C_E| \! >$ vs $\log_{2}S$ ($\Box$). The slope of the lines 
through the points is $-0.50$.

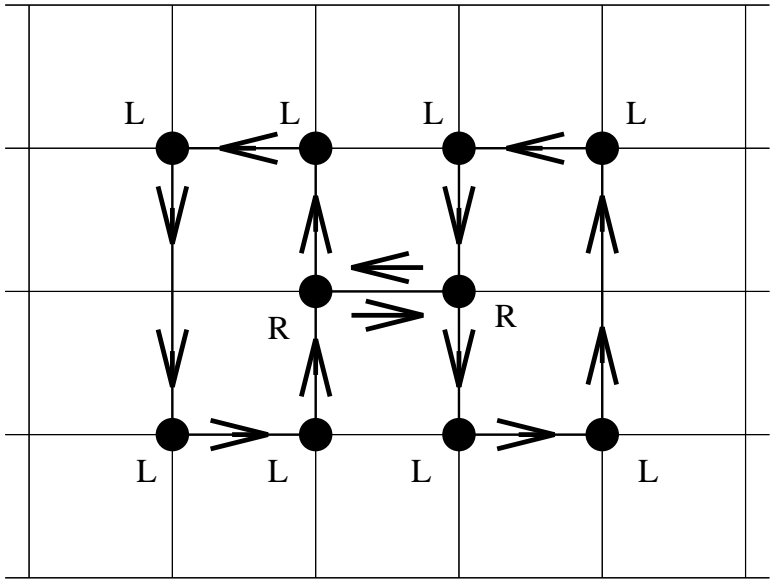
\noindent Fig. 10. The two smallest ``reflectors'', each of six 
sites with one empty site, forming a closed trajectory on the 
square lattice partially occupied by rotators. Note that neither of the
 two right rotators can be replaced by either a right mirror or 
a left mirror, so the ``reflector'' cannot be formed in the mirror 
model. We also note that this reflector is even smaller than that
 reported previously$^{(9,10)}$.

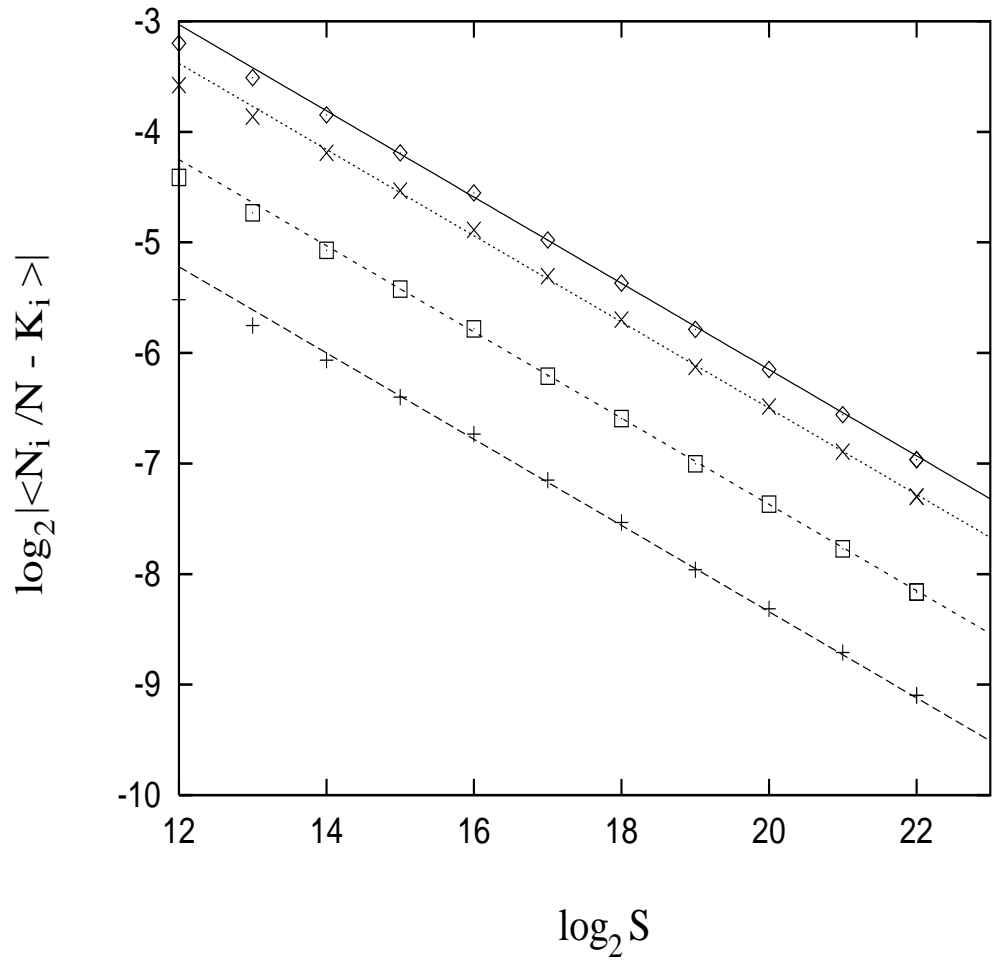
\noindent Fig. 11. Scaling behavior of $N_1$, $N_2$, $N_3$ and $
N_4$ for trajectories on the square lattice partially occupied by
 rotators, for $C=0.9$ with $C_R=0.477$, $C_L=0.423$, $C_E=0.1$
:  $\log_{2}< \! N_1/N-K_1 \! >$ vs $\log_{2}S$ ($\diamond$), 
$\log_{2}< \! N_2/N-K_2 \! >$ vs $\log_{2}S$ ($+$), $\log_{2}< \! 
K_3-N_3/N \! >$ vs $\log_{2}S$ ($\Box$), $\log_{2}< \! K_4-N_4/N
 \! >$ vs $\log_{2}S$ ($\times$), where the values of $K_1$, $K_
2$, $K_3$ and $K_4$ are $0.3047$, $0.2671$, $0.2223$ and $0.2056
$, respectively. The slope of the lines through the points is $-
0.39$.

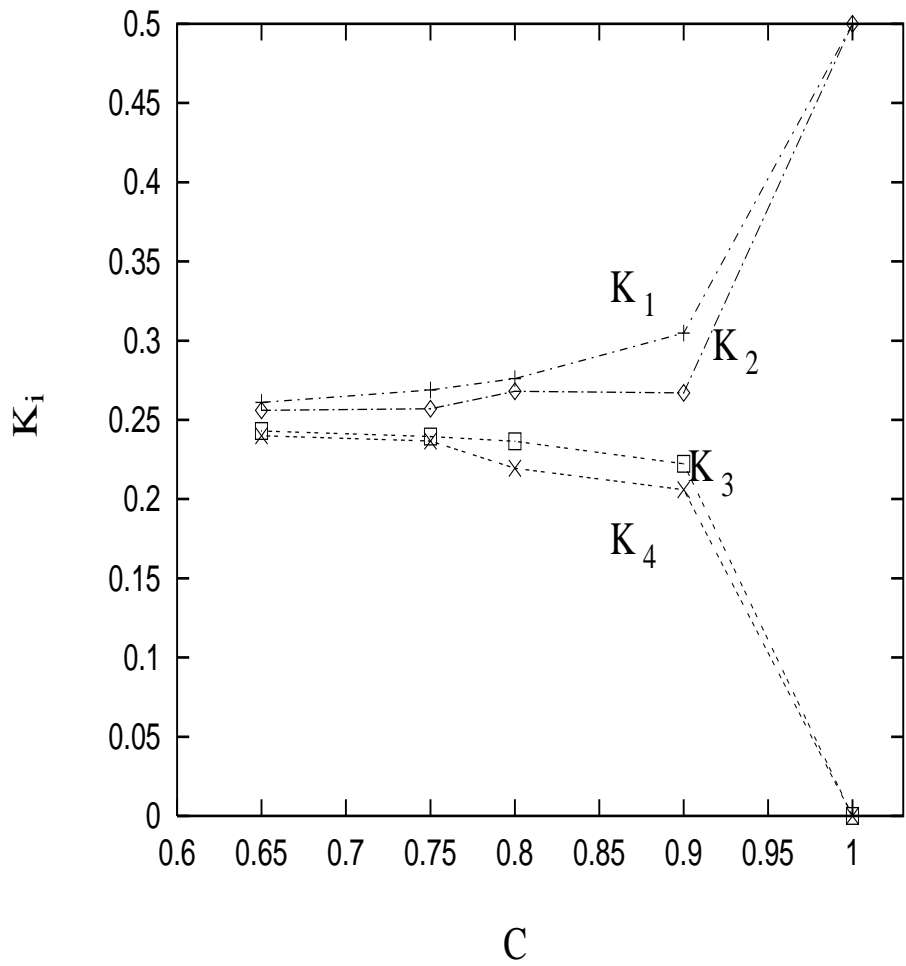
\noindent Fig. 12. Values of $K_1$ ($\diamond$), $K_2$ ($+$), $K
_3$ ($\Box$) and  $K_4$ ($\times$) for trajectories on the square
 lattice partially occupied by rotators as a function of $C$. 
The lines through the points are drawn to guide the eye.

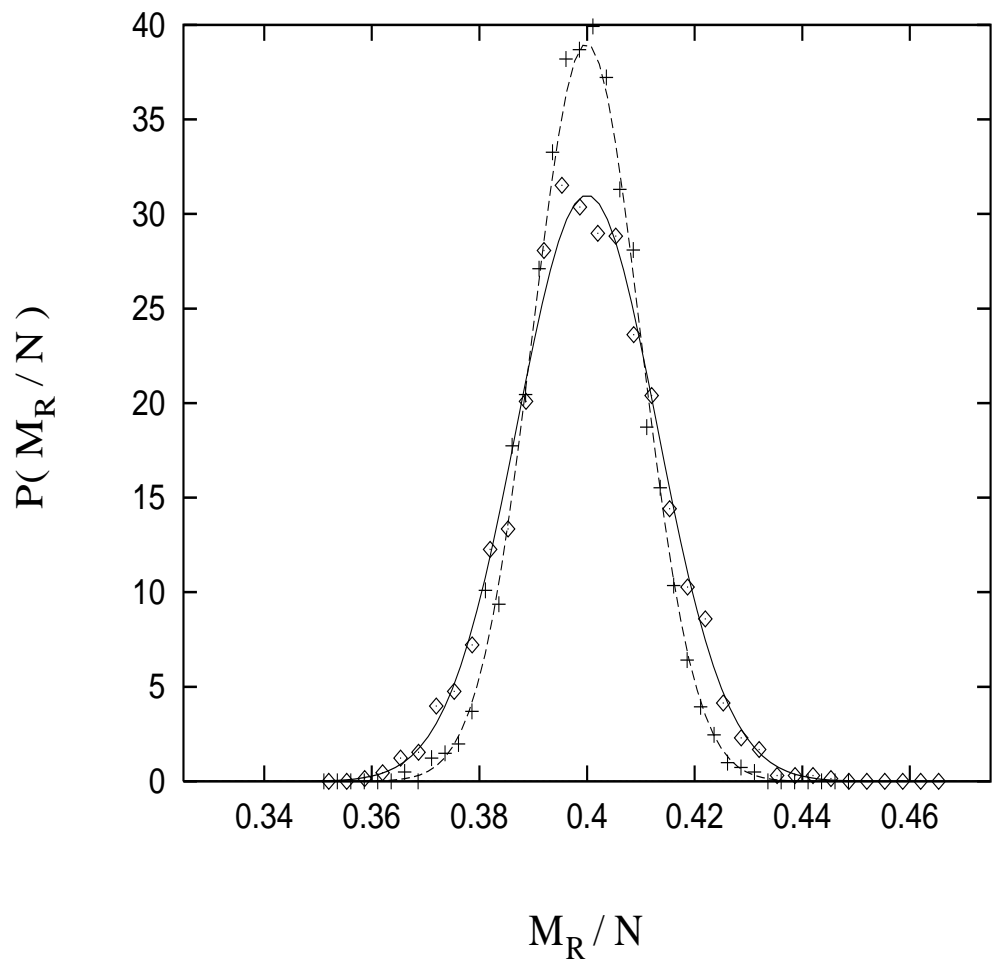
\noindent Fig. 13. The probability density $P(M_R/N)$ vs $M_R/N$
 for the mirror model on the fully occupied square lattice. The 
data were obtained from trajectories consisting of $1300$ to $15
00$ sites ($\diamond$) and $2000$ to $2300$ sites ($+$).  The solid
 and the dashed curves are described by  the Gaussians; $31 e
^{-2950(M_R/N-0.4)^2}$ and $39 e^{-4950(M_R/N-0.4)^2}$, 
respectively. 

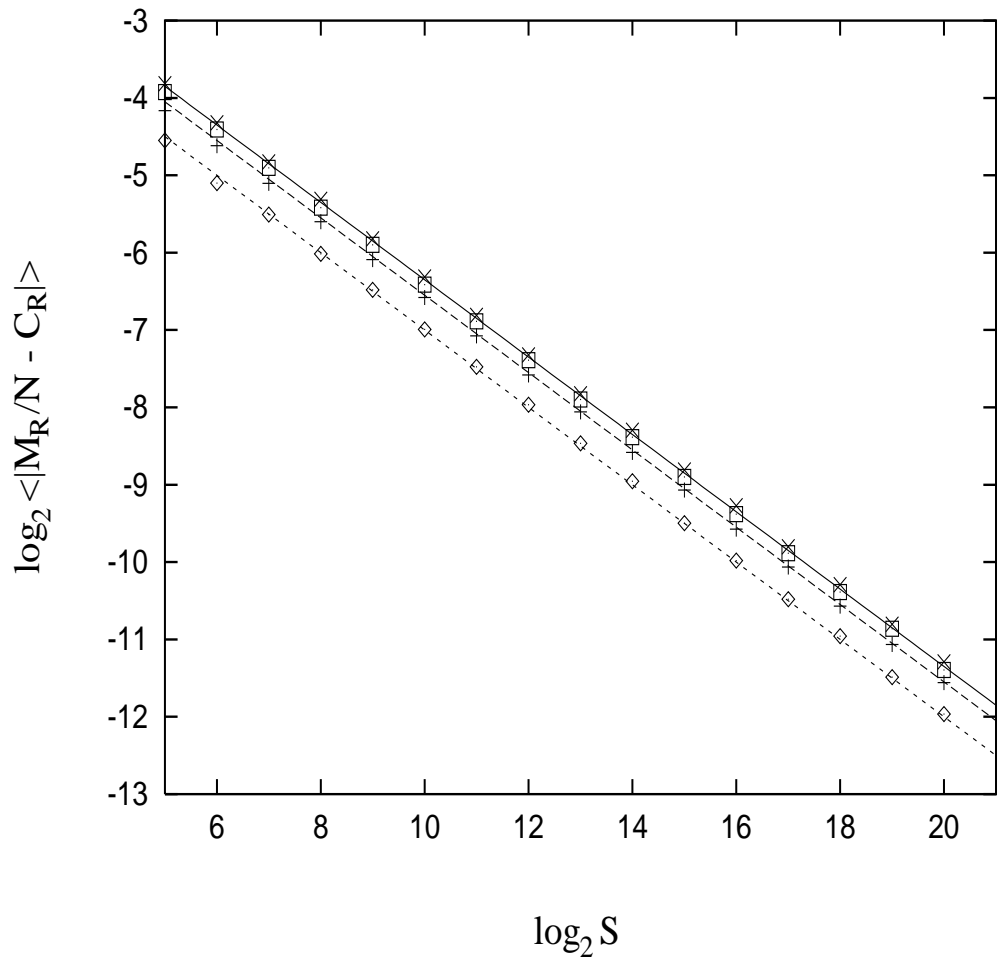
\noindent Fig. 14. Scaling behavior of $M_R$ for trajectories on
 the square lattice fully occupied by mirrors at different $C_R$
 : $\log_2<\! |M_R/N-C_R|\! >$ vs $\log_{2}S$, $C_R=0.1$ 
($\diamond$), $C_R=0.2$ ($+$), $C_R=0.3$ ($\Box$), $C_R=0.4$ ($\times$).
 The slope of the lines through the points is $-0.50$.

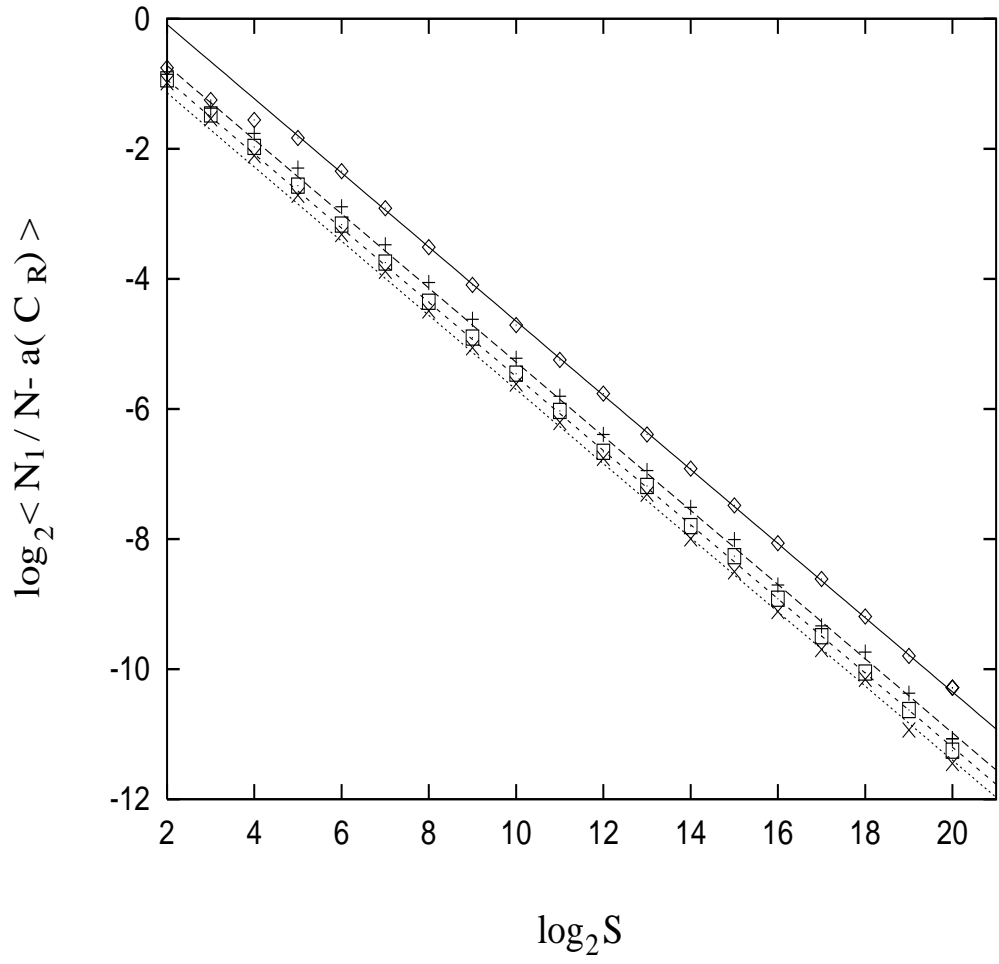
\noindent Fig. 15. Scaling behavior of $N_1$ for trajectories on
 the square lattice by mirrors at different $C_R$ along the critical
 line $C=1$ : $\log_2<N_1/N-a(C_R)>$ vs $\log_{2}S$, $C_R=0.
1$ ($\diamond$), $C_R=0.2$ ($+$), $C_R=0.3$ ($\Box$), $C_R=0.4$ 
($\times$). The slope of the lines through the points is $-0.50$.

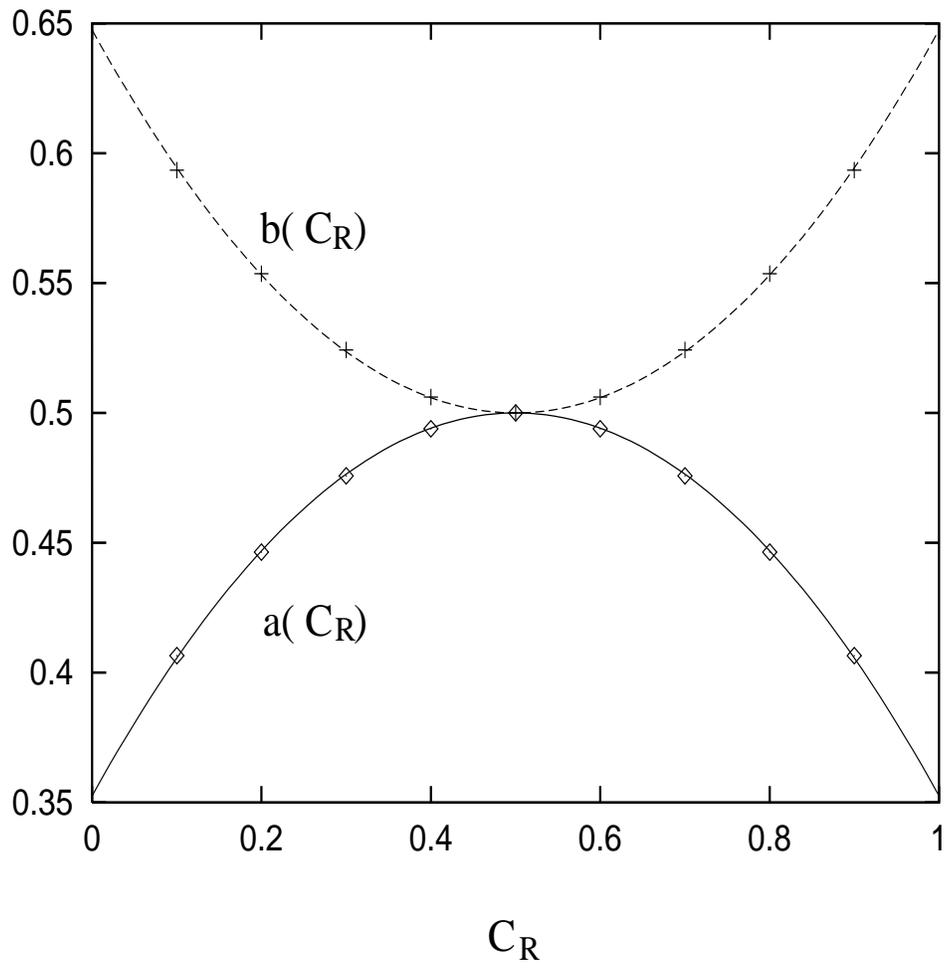
\noindent Fig. 16. For the mirror model on the fully occupied 
square lattice, $a(C_R)$ vs $C_R$ ($\diamond$), $b(C_R)$ vs $C_R$ 
($+$). The solid line and the dashed line are described by $0.5+
0.59(0.25-C_R \, C_L)$ and 
$0.5-0.59(0.25-C_R \, C_L)$, respectively.

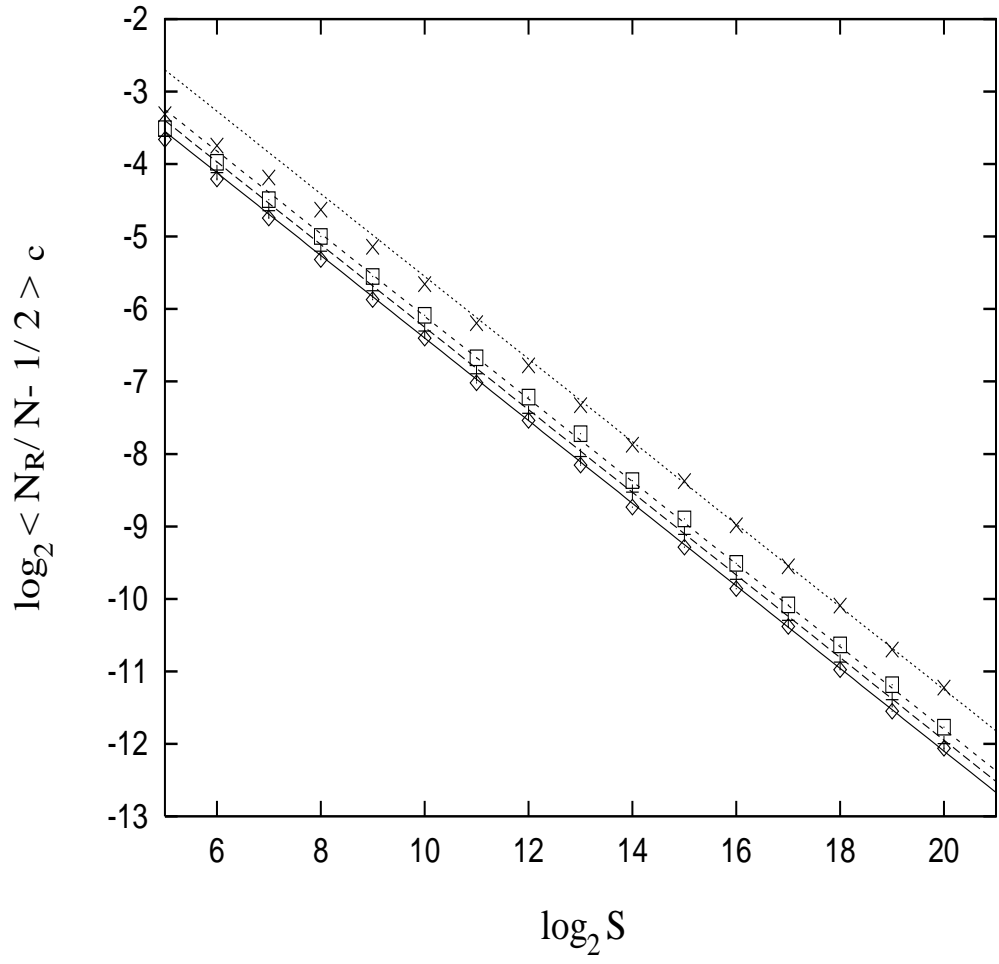
\noindent Fig. 17. Scaling behavior of $N_R$ for trajectories on
 the square lattice fully occupied by rotators mapped from mirrors,
 i.e. for the quasi-rotator model, at different $C_R$: $\log_
2< \! N_R/N-1/2 \! >_c$ vs $\log_{2}S$, $C_R=0.4$ ($\diamond$), 
$C_R=0.3$ ($+$), $C_R=0.2$ ($\Box$), $C_R=0.1$ ($\times$). The 
slope of the lines through the points is $-0.57$.

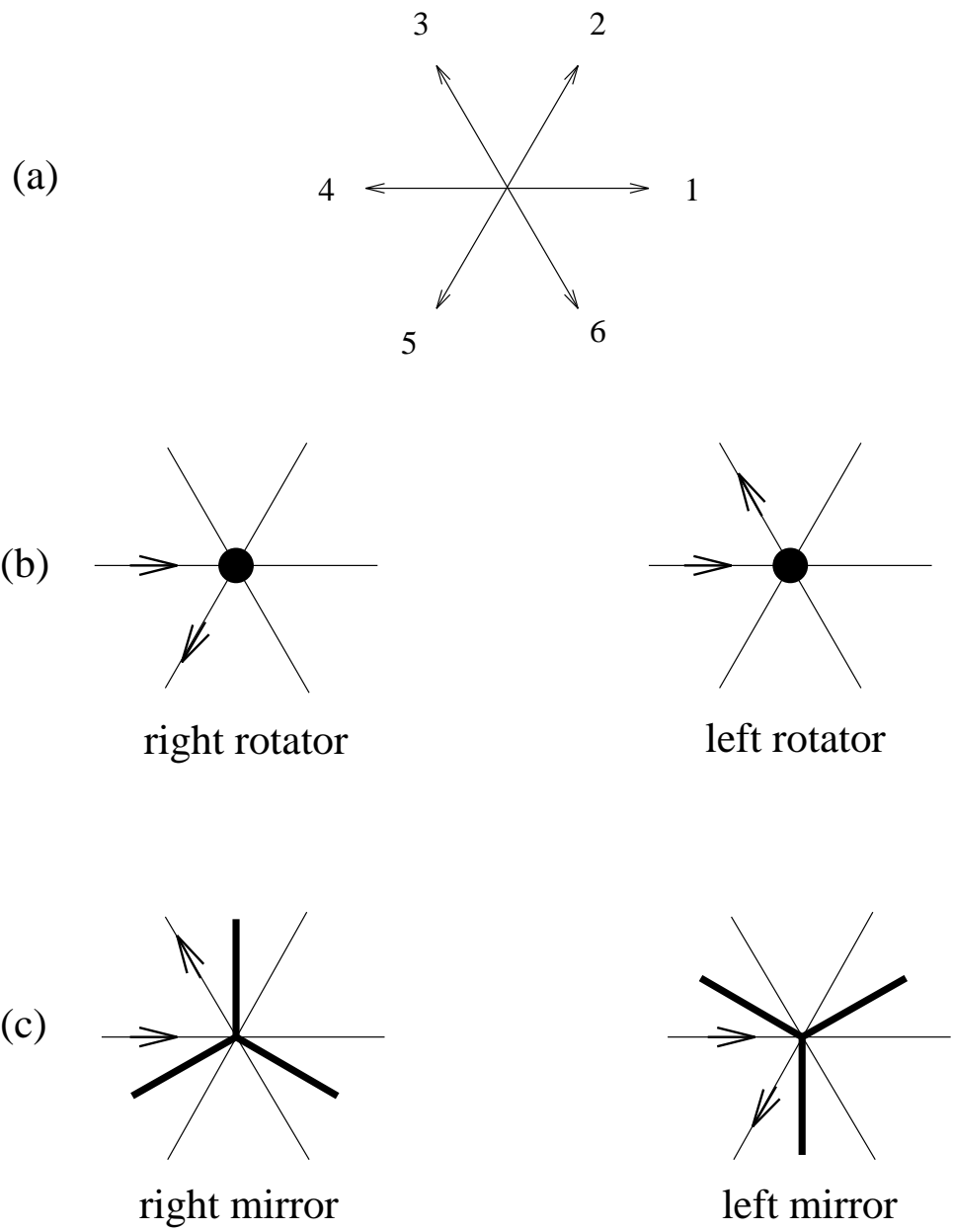
\noindent Fig. 18.  Scattering rules for the mirror and the rotator
 models on the triangular lattice. (a) the six velocity directions;
 (b) rotator scatterers; (c) mirror scatterers.

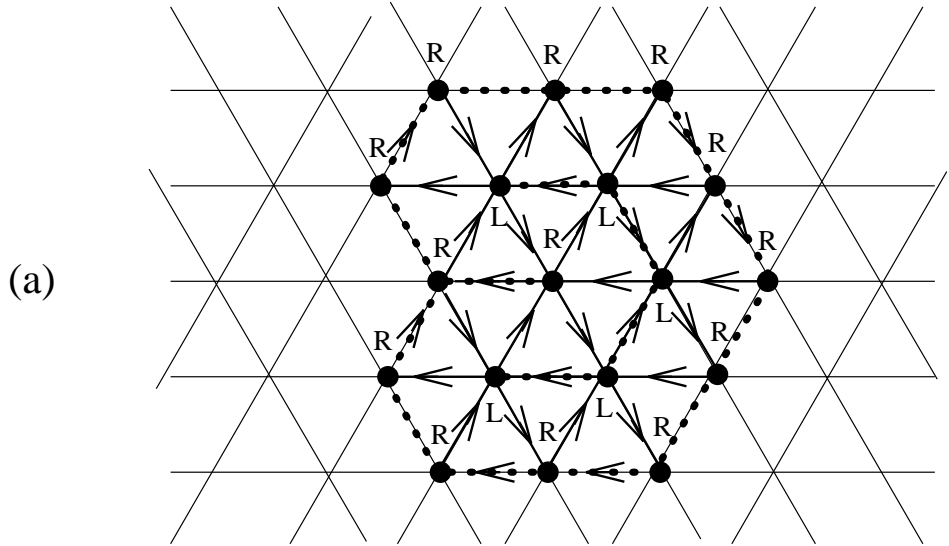
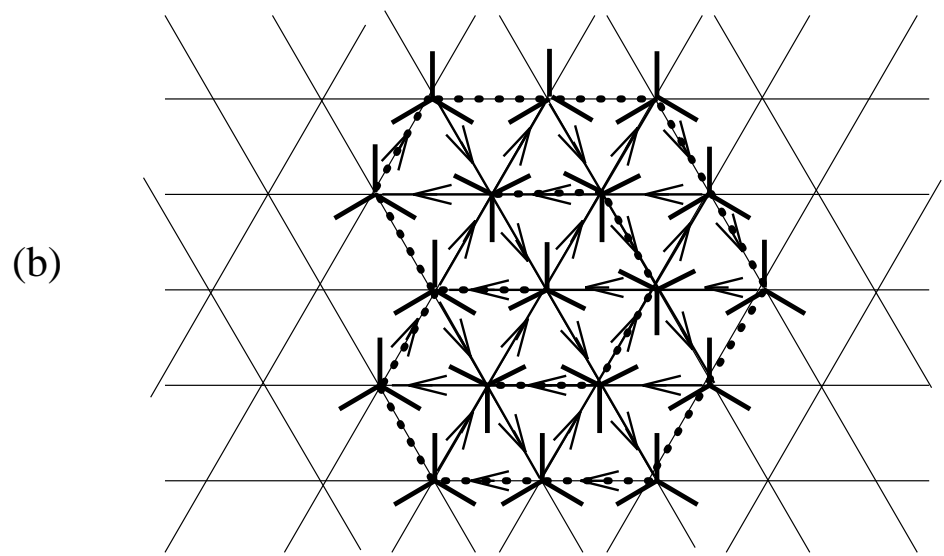
\noindent Fig. 19. (a) A typical clockwise closed trajectory on 
the triangular lattice. All right rotators and all left rotators
 are on the outer side and on the inner side of the trajectory 
respectively and form two site percolation cluster perimeters 
on the triangular lattice. Here $N_1=9$, $N_2=3$, $N_3=6$, $N_R=1
3$, $N_L=5$, $N=18$, and $S=33$. (b) the same trajectory generated
 by mirrors on the same lattice. Note all right rotators have 
been replaced by right mirrors while all left rotators have been
 replaced by left mirrors. Here $M_R=N_R=13$ and $M_L=N_L=5$.

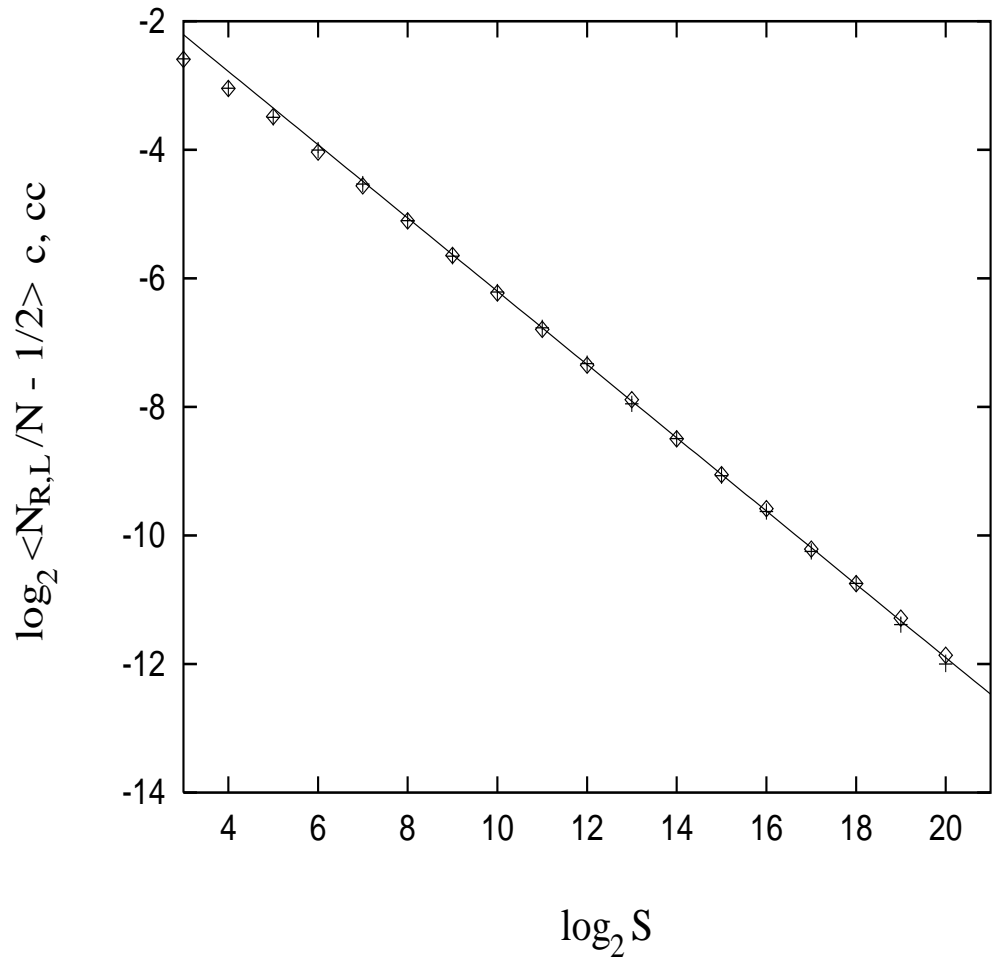
\noindent Fig. 20. Scaling behavior of $N_R$ and $N_L$ for 
trajectories on the triangular lattice fully occupied by rotators, $C
_R=C_L=1/2$:  $\log_{2}< \! N_R/N-1/2 \! >_c$ vs $\log_{2}S$ 
($\diamond$), $\log_{2}< \! N_L/N-1/2 \! >_{cc}$ vs $\log_{2}S$ ($+
$). The slope of the lines through the points is $-0.57$.

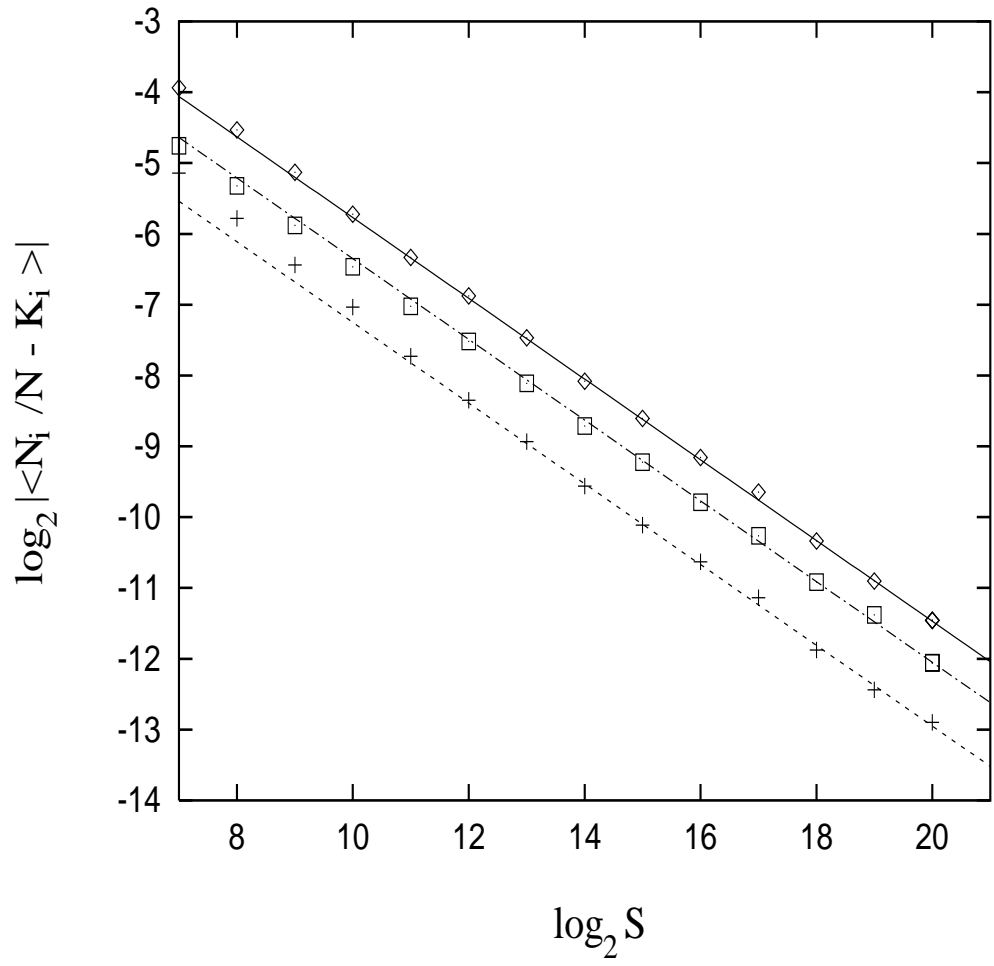
\noindent Fig. 21. Scaling behavior of $N_1$, $N_2$ and $N_3$ for
 trajectories on the triangular lattice fully occupied by rotators,
 $C_R=C_L=1/2$:  $\log_{2}< \! N_1/N-K_1 \! >$ vs $\log_{2}S
$ ($\diamond$), $\log_{2}< \! K_2-N_2/N \! >$ vs $\log_{2}S$ ($+
$). $\log_{2}< \! K_3-N_3/N \! >$ vs $\log_{2}S$ ($\Box$), where
 the values of $K_1$, $K_2$ and $K_3$ are $0.3197$, $0.4052$ and
 $0.2751$ respectively. The slope of the fitting lines are $-0.5
7$.
 
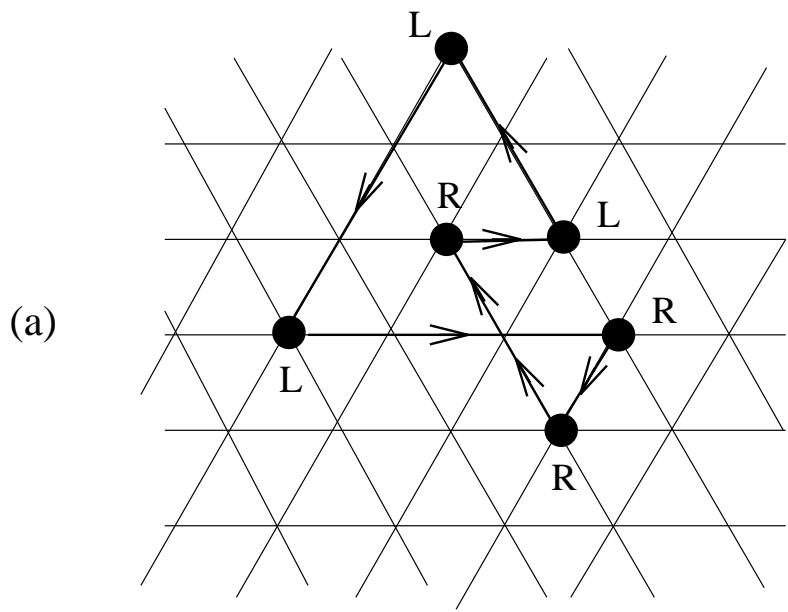
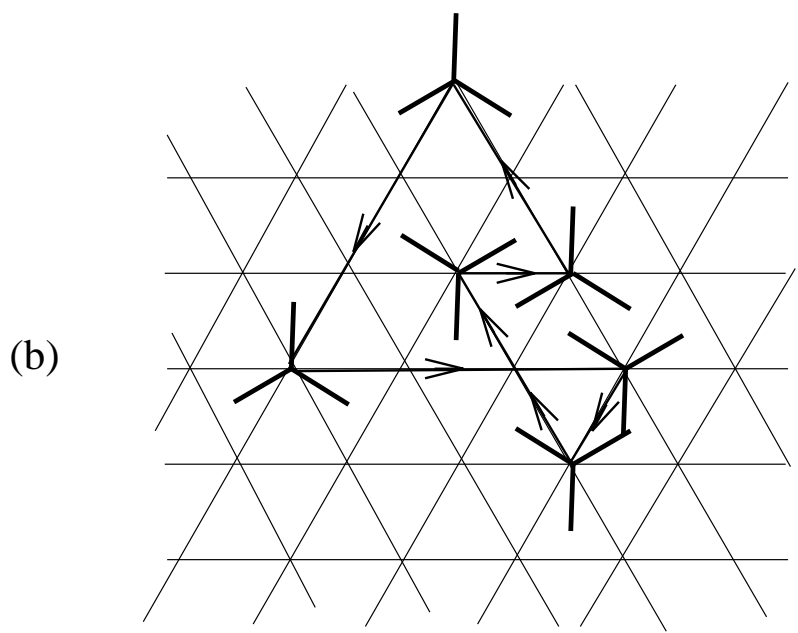
\noindent Fig. 22. (a) A typical closed trajectory generated by 
a moving particle on the triangular lattice partially occupied 
by rotators. Note that the trajectory crosses itself. Here $W=0$,
 $N_R=3$, $N_L=3$, $N_E=5$, $N_1=10$, $N_2=1$, $N_3=0$, $N=11$ a
nd $S=12$. (b) The same trajectory generated by a moving particle
 on the triangular lattice partially occupied by mirrors. Right
 rotators are replaced by right mirrors while left rotators are 
replaced by left mirrors. Here $M_R=N_R=3$ and $M_L=N_L=3$.  

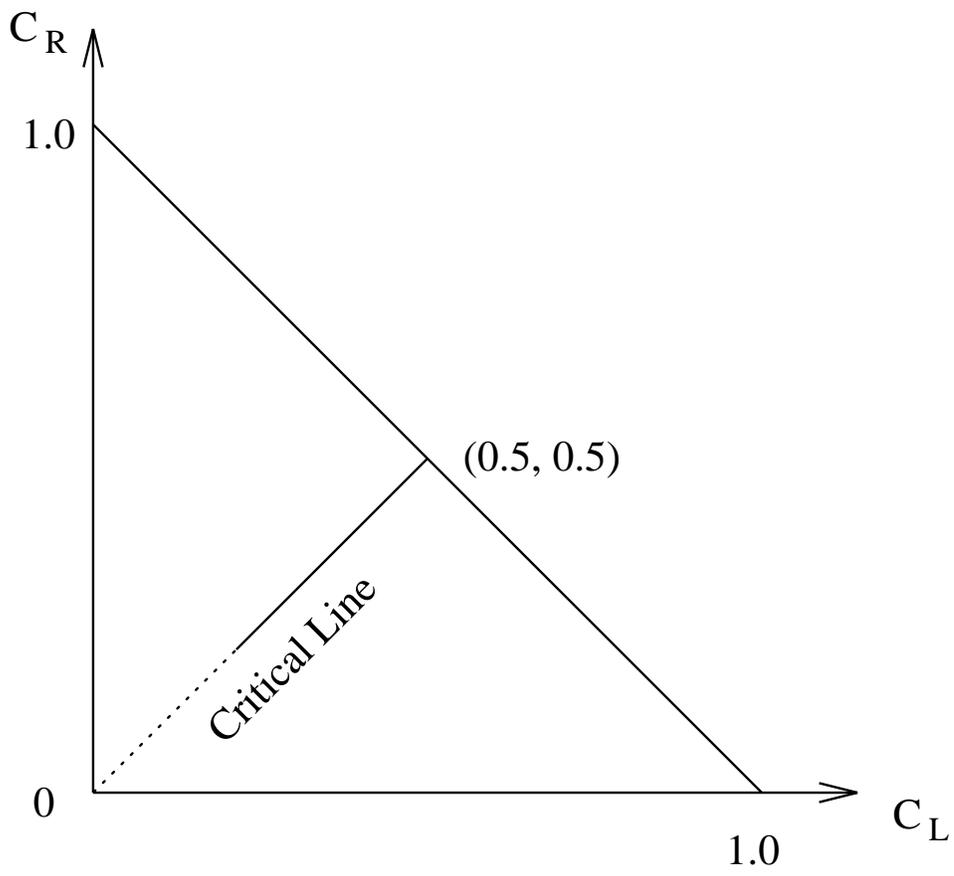
\noindent Fig. 23. Phase diagram for both the mirror and the rotator
 models on the triangular lattice. The solid critical line has
 been computed for $C>0.6$, the dashed critical line is 
conjectured.

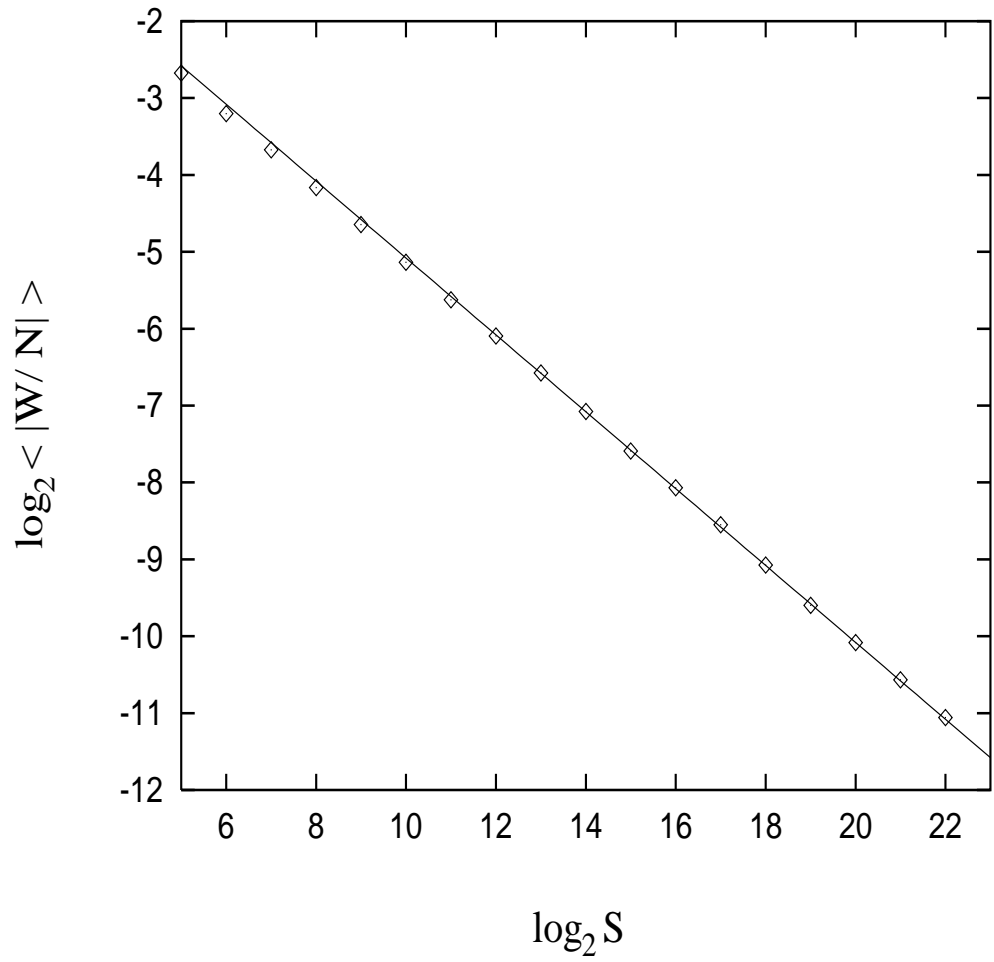
\noindent Fig. 24. Scaling behavior of $W$ for trajectories on the
 triangular lattice partially occupied by rotators, for a typical
 concentration $C=0.8$ with $C_R=C_L=0.4$: $<\! |W/N| \! >$ vs
 $\log_{2}S$ ($\diamond$). The slope of the line through the 
points is $-0.50$.

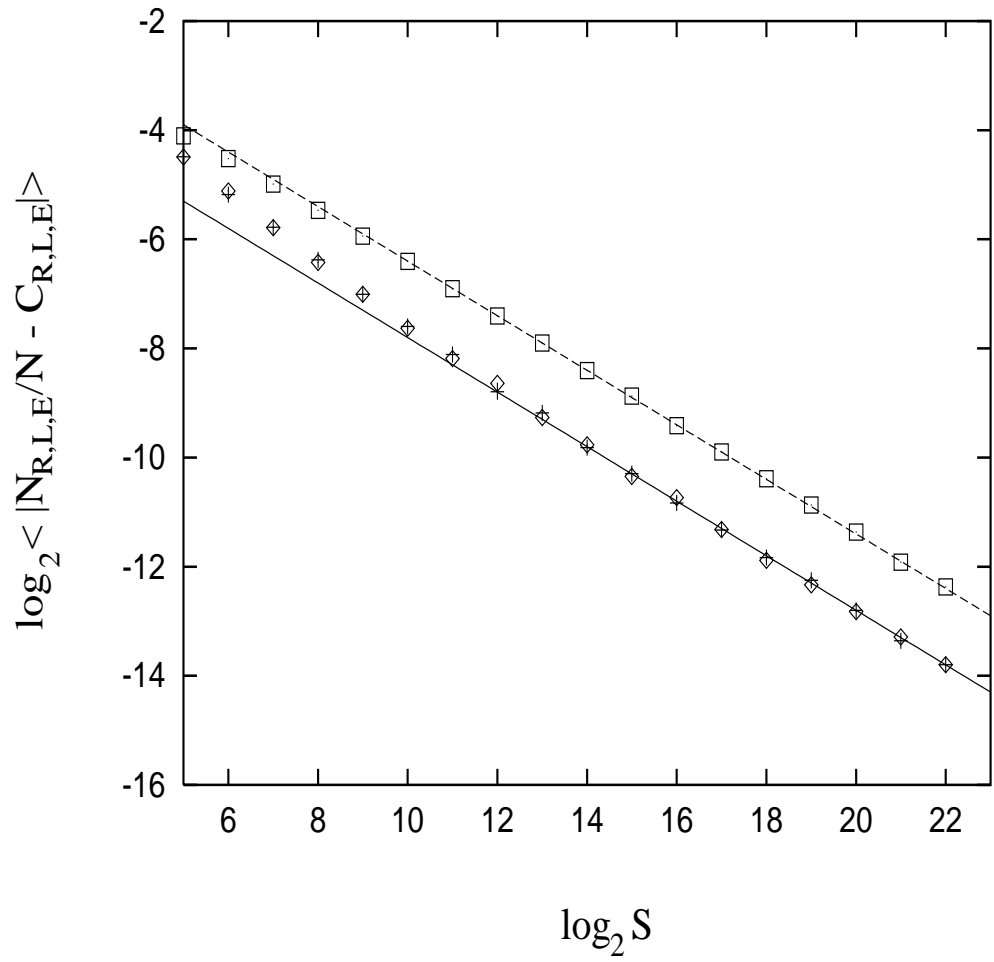
\noindent Fig. 25. Scaling behavior of $N_R$, $N_L$ and $N_E$ for
 trajectories on the triangular lattice partially occupied by 
rotators, for $C=0.8$ with $C_R=C_L=0.4$ and $C_E=0.2$:  $\log_{2
}<\! |N_R/N-C_R| \! >$ vs $\log_{2}S$ ($\diamond$), $\log_{2}
< \! |N_L/N-C_L| \! >$ vs $\log_{2}S$ ($+$), $\log_{2}<\! |N_E/N-C_
E| \! >$ vs $\log_{2}S$ ($\Box$). The slope of the lines through
 the points is $-0.50$.

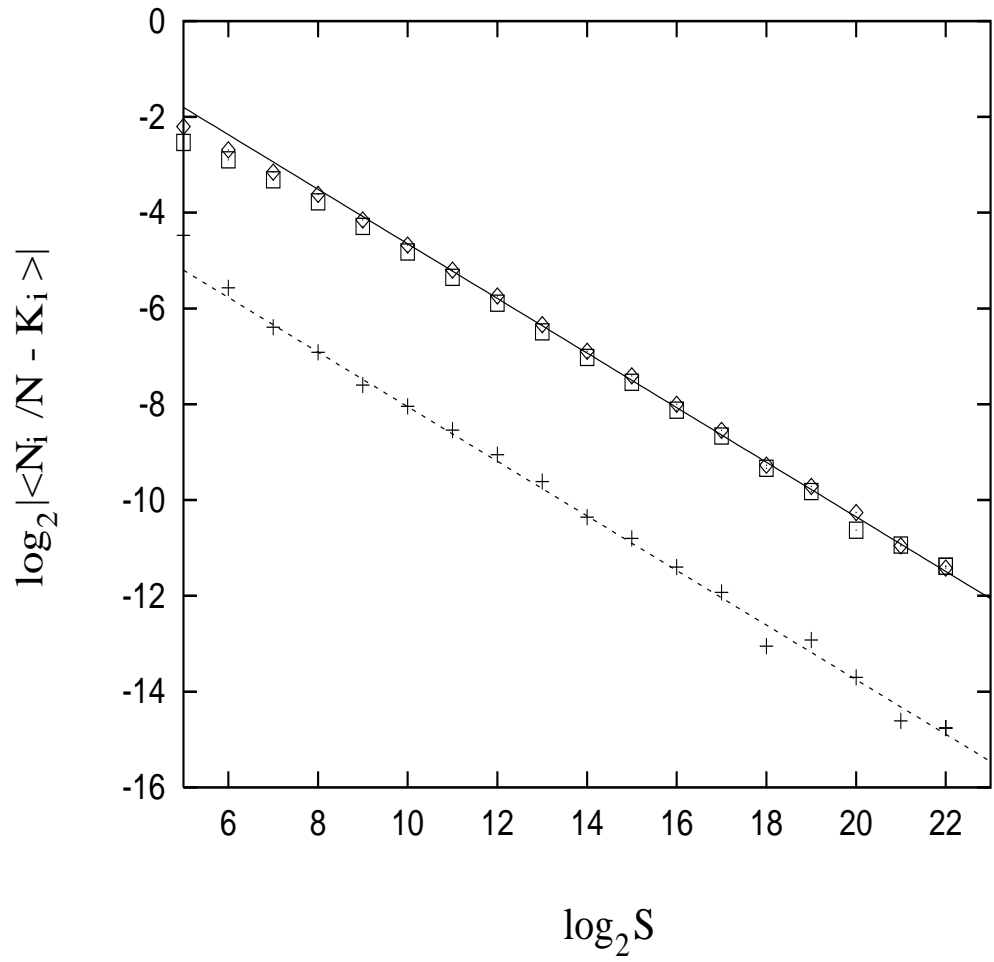
\noindent Fig. 26.  Scaling behavior of $N_1$, $N_2$ and $N_3$ 
for trajectories on the triangular lattice partially occupied by 
rotators, for $C=0.8$ with $C_R=C_L=0.4$ and $C_E=0.2$:  $\log_{
2}<\! N_1/N-K_1\! >$ vs $\log_{2}S$ ($\diamond$), $\log_{2}<\! K
_2-N_2/N \! >$ vs $\log_{2}S$ ($+$), $\log_{2}<\! K_3-N_3/N \! >
$ vs $\log_{2}S$ ($\Box$), where the values of $K_1$, $K_2$ and 
$K_3$ are $0.3356$, $0.3391$ and $0.3254$, respectively. The slope
 of the lines through the points is $-0.57$.

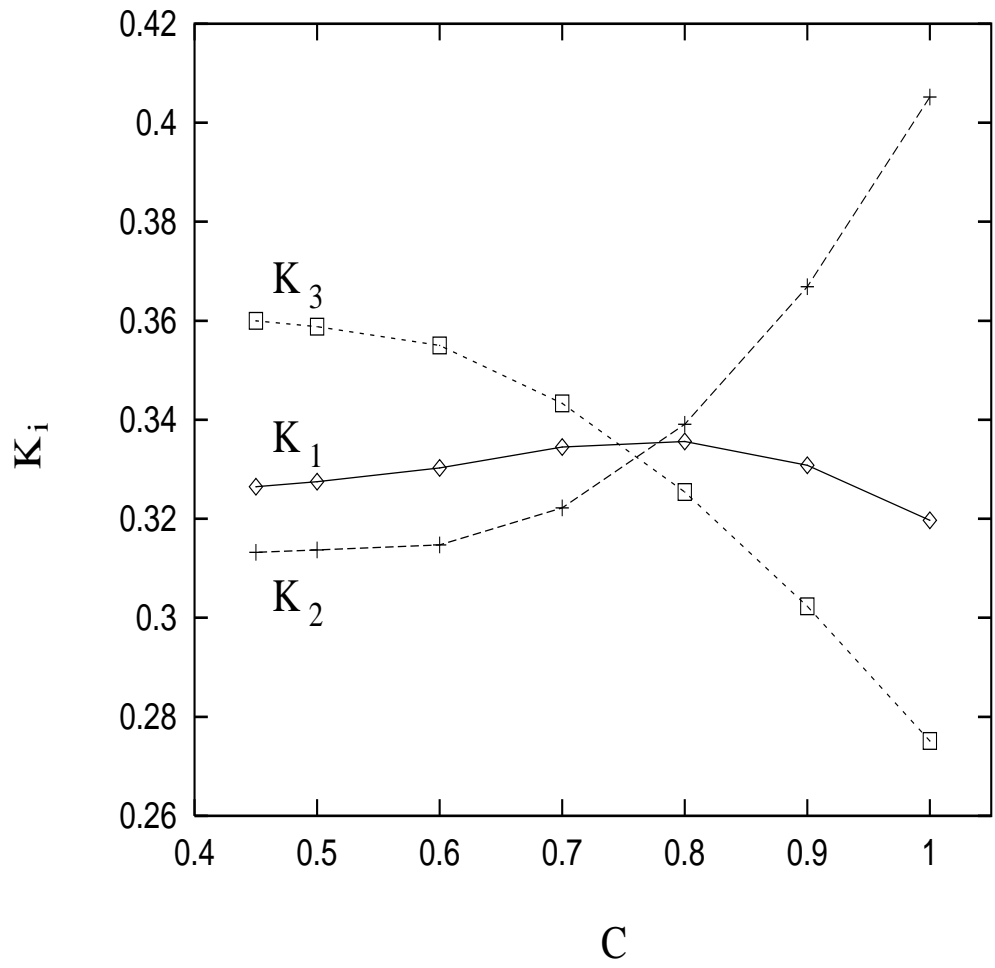
\noindent Fig. 27. Values of $K_1$ ($\diamond$), $K_2$ ($+$) and
 $K_3$ ($\Box$) for trajectories on the triangular lattice 
partially occupied by rotators as a function of $C$. The lines 
through the points are drawn to guide the eye.
\newpage
\begin{center}
{\bf{TABLE I: Critical behavior for $C=1$}}
\vskip 1.5cm
\begin{tabular}{| l| l| l|} \hline 
 \multicolumn{2}{|c|} {} & \multicolumn{1}{c|} {}  \\ 
\multicolumn{2}{|c|} {Square Lattice} & \multicolumn{1}{c|}
 {Triangular Lattice}   \\ 
 \multicolumn{2}{|c|} {} & \multicolumn{1}{c|} {}  \\ \hline

& &  \\
 \multicolumn{1}{|c|}{Rotator} & \multicolumn{1}{c|}{Mirror} & 
\multicolumn{1}{c|} {Rotator=mirror}   \\        
& & \\ \hline  \hline

& & \\
    $<\frac{ N_{^R} }{N}-\frac{1}{2}>_c=$ & $<|\frac{M_{^R}}{N}-
C_R|>=$
               &$<\frac{N_{^R}}{N}-\frac{1}{2}>_c=$ \\ 
& & \\

 $<\frac{N_{^L}}{N}-\frac{1}{2}>_{cc}\sim N^{-0.57}$
 & $<|\frac{M_{^L}}{N}-C_L|>\sim N^{-0.50}$
 & $<\frac{N_{^L}}{N}-\frac{1}{2}>_{cc}\sim N^{-0.57}$  \\  
& & \\    \hline

& & \\
  $<\frac{N_{^1}}{N}-\frac{1}{2}>=$
 &$<\frac{N_{^1}}{N}-a(C_R)>=$
 &$<\frac{N_{^1}}{N}-0.3197>\sim$    \\ 
& & \\ 
           
  $<\frac{1}{2}-\frac{N_{^2}}{N}> \sim N^{-0.57}$
 &$<b(C_R)-\frac{N_{^2}}{N}>=N^{-0.57}$
 &$<0.4052-\frac{N_{^2}}{N}>\sim$    \\ 
 & & \\
 & & $<0.2751-\frac{N_{^3}}{N}>\sim N^{-0.57}$     \\ 
& & \\ \hline

& & \\
   $<\frac{S}{N}-\frac{3}{2}> \sim N^{-0.57}$
  & $<\frac{S}{N}-1-b(C_R)>\sim N^{-0.57}$
  & $<\frac{S}{N}-1.9554> \sim N^{-0.57}$ \\        
& &  \\ \hline \hline 
\end{tabular}

\end{center}
\newpage
\begin{center}
{\bf{TABLE II: Critical behavior for $C=0.9$} }
\vskip 1.5cm
\begin{tabular}{|l| l| l| l|} \hline
 \multicolumn{2}{|c|} {} & \multicolumn{1}{c|} {}  \\ 
\multicolumn{2}{|c|} {Square Lattice} & \multicolumn{1}{c|} 
{Triangular Lattice}   \\ 
 \multicolumn{2}{|c|} {} & \multicolumn{1}{c|} {}  \\ \hline

& &  \\
 \multicolumn{1}{|c|}{Rotator} & \multicolumn{1}{c|}{Mirror} & 
\multicolumn{1}{c|} {Rotator=mirror}   \\        
& & \\ \hline  \hline

& &   \\
 $<|\frac{W}{N}-(C_R-C_L)|> \sim 2^{-2.8S^{0.18}}$ & 
               &$<|\frac{W}{N}|> \sim N^{-0.50}$ \\
& &  \\ \cline{1-1} \cline{3-3}

& &   \\
   $<|\frac{N_{^R}}{N}-C_R|>=$ & 
               &$<|\frac{N_{^R}}{N}-C_R|>=$ \\ 
& &  \\
 $<|\frac{N_{^L}}{N}-C_L|>\sim $ & 
 & $<|\frac{N_{^L}}{N}-C_L|>\sim $  \\ 

& &  \\
 $<|\frac{N_{^E}}{N}-C_E|>\sim N^{-0.50}$ 
 & Super-diffusion
 & $<|\frac{N_{^E}}{N}-C_E|>\sim N^{-0.50}$  \\

& &  \\ \cline{1-1} \cline{3-3}

& &  \\
 $<\frac{N_{^1}}{N}-0.2671>\sim$
      &   ($\tau=1$, $d_f=2$)     
      &$<\frac{N_{^1}}{N}-0.3308>\sim$    \\ 

& &  \\           
  $<\frac{N_{^2}}{N}-0.3047>\sim $ 
 & 
 &$<0.3669-\frac{N_{^2}}{N}>\sim$    \\ 

& &  \\
  $<0.2223-\frac{N_{^3}}{N}>\sim $ 
 &  
 &$<0.3023-\frac{N_{^3}}{N}>\sim N^{-0.57}$     \\ 

& & \\
 $<0.2059-\frac{N_{^4}}{N}>\sim N^{-0.39}$ & &  \\ 
& & \\  \cline{1-1} \cline{3-3} 

& &  \\
      $<\frac{S}{N}-2.3670> \sim N^{-0.39}$
      &     & $<\frac{S}{N}-1.9715> \sim N^{-0.57}$ \\    
 & &  \\ \hline \hline 
\end{tabular}

\end{center}
\newpage
\begin{center}
{\bf{TABLE III: Critical behavior for $C=0.8$} }
\vskip 1.5cm
\begin{tabular}{|l| l| l| l|} \hline
 \multicolumn{2}{|c|} {} & \multicolumn{1}{c|} {}  \\ 
\multicolumn{2}{|c|} {Square Lattice} & \multicolumn{1}{c|} 
{Triangular Lattice}   \\ 
 \multicolumn{2}{|c|} {} & \multicolumn{1}{c|} {}  \\ \hline

& &  \\
 \multicolumn{1}{|c|}{Rotator} & \multicolumn{1}{c|}{Mirror} & 
\multicolumn{1}{c|} {Rotator=mirror}   \\        
& & \\ \hline  \hline

& &   \\
 $<|\frac{W}{N}-(C_R-C_L)|> \sim 2^{-1.4S^{0.18}}$ & 
               &$<|\frac{W}{N}|> \sim N^{-0.50}$ \\
& &  \\ \cline{1-1} \cline{3-3}

& &   \\
   $<|\frac{N_{^R}}{N}-C_R|>=$ & 
               &$<|\frac{N_{^R}}{N}-C_R|>=$ \\ 
& &  \\
 $<|\frac{N_{^L}}{N}-C_L|>\sim $ & 
 & $<|\frac{N_{^L}}{N}-C_L|>\sim $  \\ 

& &  \\
 $<|\frac{N_{^E}}{N}-C_E|>\sim N^{-0.50}$ 
 & Super-diffusion
 & $<|\frac{N_{^E}}{N}-C_E|>\sim N^{-0.50}$  \\

& &  \\ \cline{1-1} \cline{3-3}

& &  \\
 $<\frac{N_{^1}}{N}-0.2680>\sim$
      &   ($\tau=1$, $d_f=2$)     
      &$<\frac{N_{^1}}{N}-0.3356>\sim$    \\ 

& &  \\           
  $<\frac{N_{^2}}{N}-0.2760>\sim $ 
 & 
 &$<0.3391-\frac{N_{^2}}{N}>\sim$    \\ 

& &  \\
  $<0.2365-\frac{N_{^3}}{N}>\sim $ 
 &  
 &$<0.3254-\frac{N_{^3}}{N}>\sim N^{-0.57}$     \\ 

& & \\
 $<0.2195-\frac{N_{^4}}{N}>\sim N^{-0.39}$ & &  \\ 
& & \\  \cline{1-1} \cline{3-3} 

& &  \\
      $<\frac{S}{N}-2.4075> \sim N^{-0.39}$
      &     & $<\frac{S}{N}-1.990> \sim N^{-0.57}$ \\    
 & &  \\ \hline \hline 
\end{tabular}

\end{center}
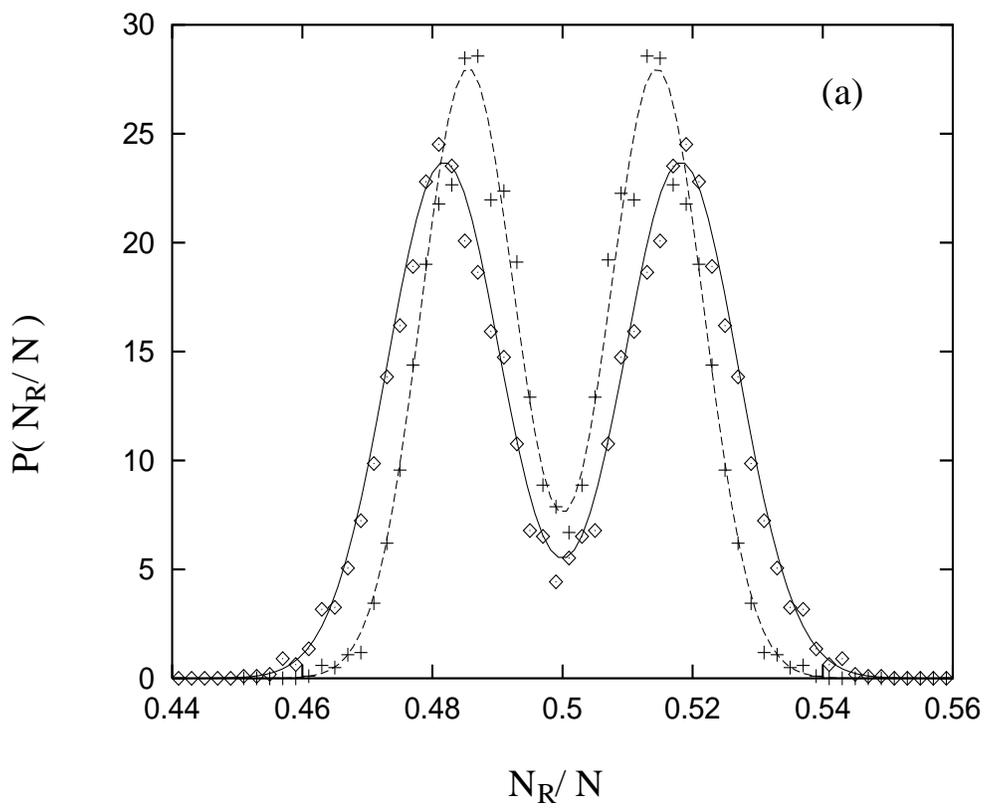
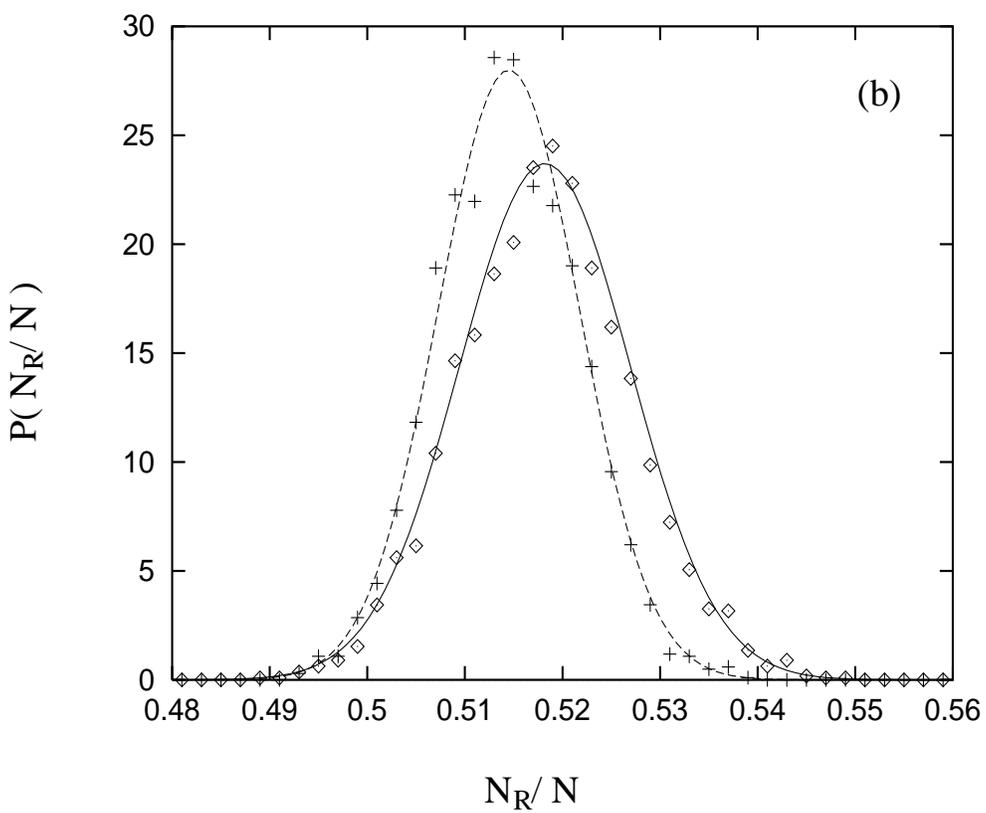
\end{document}